\documentclass[12pt,english,sort&compress]{article}
\usepackage[T1]{fontenc}
\usepackage[latin9]{inputenc}
\usepackage{fancyhdr}
\usepackage{babel}
\usepackage{booktabs}
\usepackage{amsmath}
\usepackage{graphicx}
\usepackage[a4paper]{geometry}
\usepackage{setspace}
\usepackage[numbers]{natbib}
\usepackage[pdfusetitle,
 bookmarks=true,bookmarksnumbered=false,bookmarksopen=false,
 breaklinks=false,pdfborder={0 0 0},pdfborderstyle={},backref=false,colorlinks=false]
 {hyperref}

\makeatletter

\providecommand{\tabularnewline}{\\}

\usepackage{times}

\date{}

\renewcommand{\@openbib@code}{\setlength{\itemsep}{-1pt}}

\renewcommand{\subsectionmark}[1]{}

\usepackage[compact]{titlesec}
\titleformat{\section}{\LARGE \bfseries}{\thesection}{1em}{}
\titleformat{\subsection}{\large \bfseries}{\thesubsection}{1em}{}

\makeatother

\begin{document}
\global\long\def\d{\mathrm{d}}%
\global\long\def\boltzmann{k_{b}}%

\global\long\def\entropyC{S_{c}}%
\global\long\def\energymon{u}%
\global\long\def\energyC{U_{c}}%
\global\long\def\maxfunc{\tilde{S}_{c}}%
\global\long\def\workC{W}%
\global\long\def\solidangle{\d\Gamma}%
\global\long\def\partition{Z}%
\global\long\def\pdf{p_{\theta}}%

\global\long\def\force{f}%
\global\long\def\etoe{r}%
\global\long\def\etoeTdir{\hat{\mathbf{r}}}%
\global\long\def\monomers{n}%
\global\long\def\lenref{l_{0}}%
\global\long\def\lencur{l}%

\global\long\def\pbond{p_{b}}%
\global\long\def\pchain{p_{c}}%
\global\long\def\Eactivation{E_{a}}%
\global\long\def\leneq{\lencur_{eq}}%
\global\long\def\leneqmax{\lencur_{TS}}%
\global\long\def\De{D_{e}}%

\global\long\def\forcemon{\force_{b}}%
\global\long\def\forcemax{\force_{m}}%
\global\long\def\frat{\tau}%
\global\long\def\bonds{m}%

\global\long\def\uTS{u_{TS}}%
\global\long\def\ueq{u_{eq}}%

\global\long\def\rat{\rho}%

\global\long\def\stressT{\boldsymbol{\sigma}}%
\global\long\def\piolaT{\mathbf{P}}%

\title{\vspace{-80pt}Rethinking failure in polymer networks: a probabilistic
view on progressive damage}
\author{Noy Cohen$^{\footnotesize{a,b,}} \footnote{e-mail address: noyco@technion.ac.il}\,\,$,
Nikolaos Bouklas$^{\footnotesize{b,c}}\,\,$, and Chung-Yuen Hui$^{\footnotesize{b}}\,\,$
\\
$^{a}${\footnotesize{Department of Materials Science and Engineering, Technion  - Israel Institute of Technology, Haifa 3200003, Israel}}\\
$^{b}${\footnotesize{Sibley School of Mechanical and Aerospace Engineering, Cornell University, Ithaca, New York 14853, USA}}\\
$^{c}${\footnotesize{Pasteur Labs, Brooklyn, NY 11205, USA}}}
\maketitle
\begin{abstract}
The mechanics of single-chain stretching and rupture are central to
understanding the resilience of biological polymers and designing
strong and tough soft materials such as double-network gels and multi-network
elastomers. In this work, we develop a statistical mechanics based
model that enables one to determine the distribution of forces along
the chain segments. By combining the force distribution with a tilted
bond potential that captures the stretch energy stored in these bonds,
we calculate the corresponding activation energy required for bond
dissociation. This allows us to determine the probability of bond
(and consequently chain) failure. The proposed approach is simple,
direct, and readily adaptable for constructing higher-level coarse-grained
descriptions of damage and fracture in polymer networks. We demonstrate
this by applying the theory to two problems of practical interest:
(1) toughening networks via sacrificial bond rupture in polymer chains
and (2) incorporation of the local chain model into a 3-dimensional
constitutive relation that captures damage in elastomers. The latter
was implemented through the micro-sphere framework, which accounts
for different chain orientations, as well as the computationally inexpensive
eight chain model. The findings from this work provide a physically-based
model to quantify the stretching and failure of a single chain and
pave the way to the integration of local damage models into 3-dimensional
networks.
\end{abstract}


\section{Introduction}

The stretching and rupture of individual polymer chains underpin modern
multi-scale descriptions of damage, fracture, and adhesion in soft
polymeric materials \citep{Ghatak2000,Lavoie2016,Mao2017,Lamont2021,Lamont2025,Wang2025}.
Constitutive models such as the three-, the four-, and the eight-chain
network models \citep{Wang1952,Treloar1946,Arruda1993} describe the
nonlinear elasticity of rubbery networks comprising freely jointed
chains (FJCs) \citep{Kuhn1942,Zhu2025}. When chain rupture is incorporated,
these models naturally extend to continuum descriptions of damage
and Mullins-type softening \citep{Webber2007,Wang2011}, readily providing
a molecularly informed theory for adhesion models in elastomers \citep{Brown1994,Creton2002,Lavoie2019,Yang2020}.

Single-chain mechanics also appear explicitly in theories of fracture
and adhesion. The classical Lake--Thomas (LT) theory relates the
energy required to propagate a crack to the energy needed to rupture
a single polymer chain between cross-link junctions, making chain-level
failure the fundamental dissipative mechanism governing toughness
\citep{Lake1967}. This mechanism---energy dissipation through chain
stretching and scission---has since been recognized as a central
ingredient enabling the design of exceptionally tough soft materials
such as double-network hydrogels \citep{Li2024,Gong2003,Gong2010}
and multi-network elastomers \citep{Ducrot2014}, where controlled
chain damage in sacrificial networks provides large distributed dissipation
during crack growth. Recognizing that LT is a scaling law stemming
from the idealization that rupture takes place on a single plane,
recent works have also focused on the limitations of this assumption
towards a more realistic resolution of damage zone size \citep{Slootman2020}. 

Many works developed models that account for rupture through probabilistic
considerations \citep{Vernerey2018,Itskov2016,Itskov2016a}, limiting
values (in which failure is accounted for phenomenologically) \citep{Mao2017,Mulderrig2021,Volokh2010,Volokh2013},
and asymptotic approximations and matching \citep{Mulderrig2023,Buche2022,Buche2020,Buche2021}.
Similar concepts guide models of protein-based networks \citep{Olive2024,Hillgaertner2018,Cohen2025,Du2011}
and adhesive failure, where debonding proceeds through sequential
stretching and detachment of load-bearing chains or chain-like bridges
\citep{Qian2017,Keren2023}. 

Recent advances in single-molecule force spectroscopy now allow direct
measurement of the energy stored in an individual polymer chain at
rupture, enabling quantitative comparisons with the LT hypothesis
that the chain-breaking energy equals the number of bonds in a chain
times the C--C bond dissociation energy \citep{Zhang&etal00JPCB,Abkenar2017,Zhang2003,Beyer2000}.
A recent analysis by \citet{Wang2019} employed single-molecule force
spectroscopy data to demonstrate that the stored energy per bond at
the onset of fracture is approximately one-third of the dissociation
energy of a C--C bond. Thus, the LT theory underestimates the fracture
energy by about 60\%. More advanced models for fracture have borrowed
ideas from the phase field \citep{Miehe2014,Ang2022} and gradient
enhanced \citep{PEERLINGS1996,Mousavi2024} models, merging these
with statistical mechanics theories \citep{Talamini2018,Mousavi2025,Mousavi2026}.

In this contribution, statistical mechanics methodology is applied
to better understand the origin of failure in chains. To this end,
we derive a framework that approximates the distribution of forces
and the consequent enthalpy of repeat units along a chain, an aspect
that has been neglected in previous theories. We show that the forces
are not distributed equally among chain segments, and the force is
proportional to the orientation of a segment with respect to the end-to-end
vector. By employing the tilted potential energy profile for the bonds
along the chain \citep{Kauzmann1940,Evans1997}, we compute the activation
energy required to dissociate a bond. This energy decreases as the
chain stretches, and accordingly the probability of rupture increases.
This approach is advantageous since it accounts for the stochastic
nature of chain scission and inherently yields a maximum stretching
force that can be exerted on the chain, above which bonds break spontaneously. 

To demonstrate the predictive capability of the proposed theory, we
apply it to two representative problems of practical interest. The
first concerns chains with sacrificial bonds, in which sequential
bond rupture governs energy dissipation and toughening. This is a
common occurrence in biological polymers \citep{Fantner2006,Evans1997}.
 The second application incorporates the chain model into a continuum
network representation through (1) a micro-sphere-based integration
\citep{Bazant&oh86ZAMM,Miehe2004,Mulderrig2021} and (2) the eight
chain model of \citet{Arruda1993}. These frameworks are employed
to study anisotropic damage accumulation under large deformations.
This model can be naturally embedded into phase field and gradient
damage frameworks, opening a new avenue for future studies of fracture
in elastomers.

\section{The free energy of a chain}

Consider a FJC comprising $\monomers$ stretchable Kuhn segments with
an initial (relaxed) length $\lenref$. The chain is and its Kuhn
segments stretch such that the end-to-end distance and direction are
$\etoe$ and $\etoeTdir$, respectively. We point out that one can
either prescribe the stretching force $\force$ or the end-to-end
distance $\etoe$ along the direction $\etoeTdir$.

The number of conformations available to the chain under the end-to-end
constraint is $\Omega=\monomers!/\left(\Pi_{i}\monomers^{\left(i\right)}!\right)$,
where $\monomers^{\left(i\right)}$ is the number of Kuhn segments
that form an angle $\theta^{\left(i\right)}\le\theta\le\theta^{\left(i\right)}+\d\theta$
with the end-to-end vector $\etoeTdir$ \citep{Flory53book,Treloar75book}.
We denote the deformed length of the $i$-th segment by $\lencur^{\left(i\right)}$. 

The entropy of the chain is
\begin{equation}
\entropyC=\boltzmann\ln\left(\Omega\right)\approx\boltzmann\left(\monomers\ln\monomers-\monomers-\sum_{i}\left(\monomers^{\left(i\right)}\ln\left(\monomers^{\left(i\right)}\right)-\monomers^{\left(i\right)}\right)\right),\label{eq:entropy}
\end{equation}
where Stirling's approximation is employed and $\boltzmann$ is the
Boltzmann's constant. The chain is subjected to the constraints 
\begin{align}
\monomers & =\sum_{i}\monomers^{\left(i\right)},\label{eq:constraint1}\\
\etoe & =\sum_{i}\lencur^{\left(i\right)}\monomers^{\left(i\right)}\cos\theta^{\left(i\right)},\label{eq:constraint2}\\
\energyC & =\sum_{i}\monomers^{\left(i\right)}\energymon^{\left(i\right)},\label{eq:constraint3}
\end{align}
where $\energymon^{\left(i\right)}=\energymon\left(\lencur^{\left(i\right)}\right)$
is the potential energy of the $i$-th segment due to stretching and
$\energyC$ is the total energy of the chain.

Following common practice \citep{Flory53book,Treloar75book}, we assume
that the chain occupies the most probable conformation under the given
constraints. The entropy maximization description for bond angles
and segment length can be justified as follows: at low chain extensions,
the conformational distribution of the segments is wide and therefore
the chain fluctuates significantly. Consequently, the entropic cost
of stretching the chain through the rotation of the freely jointed
segments is significantly smaller than the cost of extending them.
Therefore, this framework directly follows the common assumptions
of \citet{Flory53book,Treloar75book}. As the chain stretches, the
conformational distribution becomes asymptotically concentrated and
the variance in the conformational landscape decreases. In this case,
the energetic cost of extending the segments is lower than the entropic
cost of stretching the chain through segment rotation. 

Taking the most probable conformation is equivalent to maximizing
the {\small entropy under the given constraints, i.e. 
\begin{equation}
\maxfunc\left(\monomers^{\left(i\right)},\lencur^{\left(i\right)},\alpha,\beta,\gamma\right)=\entropyC+\boltzmann\left(\alpha\left(\sum_{i}\monomers^{\left(i\right)}-\monomers\right)+\frac{\beta}{\lenref}\left(\sum_{i}\lencur^{\left(i\right)}\monomers^{\left(i\right)}\cos\theta^{\left(i\right)}-\etoe\right)+\gamma\left(\sum_{i}\monomers^{\left(i\right)}\energymon^{\left(i\right)}-\energyC\right)\right),\label{eq:function_to_maximize}
\end{equation}
}where $\alpha$, $\beta$, and $\gamma$ are Lagrange multipliers
that enforce the constraints in Eqs. \ref{eq:constraint1}, \ref{eq:constraint2},
and \ref{eq:constraint3}. The maximization of Eq. \ref{eq:function_to_maximize}
with respect to $\monomers^{\left(i\right)}$ and $\lencur^{\left(i\right)}$
yields
\begin{equation}
\frac{\partial\maxfunc}{\partial\monomers^{\left(i\right)}}=\boltzmann\left(-\ln\left(\monomers^{\left(i\right)}\right)+\alpha+\beta\frac{\lencur^{\left(i\right)}}{\lenref}\cos\theta^{\left(i\right)}+\gamma\,\energymon^{\left(i\right)}\right)=0,\label{eq:differential_n_equation}
\end{equation}
and 
\begin{equation}
\frac{\partial\maxfunc}{\partial\lencur^{\left(i\right)}}=\boltzmann\monomers^{\left(i\right)}\left(\frac{\beta}{\lenref}\cos\theta^{\left(i\right)}+\gamma\,\frac{\partial\energymon^{\left(i\right)}}{\partial\lencur^{\left(i\right)}}\right)=0,\label{eq:differential_l_equation}
\end{equation}
respectively. From Eqs. \ref{eq:differential_n_equation} and \ref{eq:differential_l_equation},
one can obtain
\begin{equation}
\monomers^{\left(i\right)}=\exp\left(\alpha\right)\exp\left(\beta\frac{\lencur^{\left(i\right)}}{\lenref}\cos\theta^{\left(i\right)}+\gamma\,\energymon^{\left(i\right)}\right),\label{eq:segment_distribution}
\end{equation}
and 
\begin{equation}
\frac{\partial\energymon^{\left(i\right)}}{\partial\lencur^{\left(i\right)}}=-\frac{\beta}{\lenref\gamma}\cos\theta^{\left(i\right)}.\label{eq:equation_for_bond_length}
\end{equation}
Note that as opposed to previous works \citep{Buche2022,Mulderrig2023},
the current formulation can also be used to determine the deformed
length of a Kuhn segment corresponding to the maximum entropy configuration
under prescribed boundary conditions. 

Substituting Eq. \ref{eq:segment_distribution} into the Eq. \ref{eq:entropy}
yields the entropy
\begin{equation}
\entropyC=\boltzmann\left(\monomers\ln\monomers-\alpha\monomers-\frac{\beta}{\lenref}\,\etoe-\gamma\,\energyC\right).\label{eq:entropy_C}
\end{equation}

It remains to determine the Lagrange multipliers. The first law of
thermodynamics reads $\d\energyC=\d\workC+T\d\entropyC$, where $\d\workC$
is an increment in the work done on the system. Accordingly, one can
write 
\begin{equation}
\frac{\partial\energyC}{\partial\energyC}=T\frac{\partial\entropyC}{\partial\energyC}\Rightarrow\gamma=-\frac{1}{\boltzmann T},
\end{equation}
which yields
\begin{equation}
\energyC-T\entropyC=-\boltzmann T\left(\monomers\ln\monomers-\alpha\monomers-\frac{\beta}{\lenref}\,\etoe\right).
\end{equation}

Next, we employ Eq. \ref{eq:segment_distribution} along with the
constraint in Eq. \ref{eq:constraint1} to write $\exp\alpha=\monomers/\partition$,
where
\begin{equation}
\partition=\int_{0}^{\pi}\exp\left(\beta\,\frac{\lencur}{\lenref}\cos\theta-\frac{\energymon\left(\lencur\right)}{\boltzmann T}\right)\solidangle,\label{eq:partition}
\end{equation}
is the partition function and $\solidangle=\sin\theta\,\d\theta/2$
is the solid angle-based probability density function with $0\le\theta\le\pi$.
Note in Eq. \ref{eq:partition} we convert the discrete summation
to an integral and accordingly the deformed Kuhn length $\lencur=\lencur\left(\theta,\beta\left(\etoe\right)\right)$. 

With Eqs. \ref{eq:segment_distribution} and \ref{eq:partition},
we define the probability that a segment forms an angle $\theta$
with $\etoeTdir$,
\begin{equation}
\pdf\left(\theta,\beta\left(\etoe\right)\right)=\frac{\monomers^{\left(i\right)}}{\monomers}=\frac{1}{\partition}\exp\left(\beta\,\frac{\lencur}{\lenref}\cos\theta-\frac{\energymon}{\boltzmann T}\right),\label{eq:probability}
\end{equation}
such that $\int_{0}^{\pi}\pdf\solidangle=1$. Using Eq. \ref{eq:probability},
Eq. \ref{eq:constraint2} is 
\begin{equation}
\rat=\frac{\etoe}{\monomers\lenref}=\int_{0}^{\pi}\frac{\lencur}{\lenref}\cos\theta\pdf\left(\theta,\beta\left(\etoe\right)\right)\solidangle,\label{eq:end-to-end_equation}
\end{equation}
where we define the ratio $\rat$ between the end-to-end distance
and the initial contour length of the chain and employ the relation
$\monomers^{\left(i\right)}=\monomers\pdf$. 

The force on the chain can be computed via (see derivation in Appendix
\ref{sec:Derivation-of-force_on_chain})
\begin{equation}
\force=\frac{\partial\workC}{\partial\etoe}=\frac{\partial\left(\energyC-T\entropyC\right)}{\partial\etoe}=\frac{\boltzmann T}{\lenref}\beta.\label{eq:force_chain}
\end{equation}

Before proceeding, we point out that the above framework recovers
the classical FJC model in the case of inextensible bonds (i.e. $\energymon=0$).
The incorporation of an enthalpic term enables to capture the distribution
of forces along the chain. Specifically, Eqs. \ref{eq:equation_for_bond_length}
and \ref{eq:force_chain} can be combined to give 
\begin{equation}
\frac{\partial\energymon}{\partial\lencur}=\forcemon,\label{eq:force_on_bond}
\end{equation}
where $\forcemon=\force\cos\theta$ is the force on a Kuhn segment.
This relation reveals that segments along the end-to-end vector experience
different forces, with a maximum of $\forcemon=\force$ for segments
that are aligned along $\etoeTdir$ (i.e. $\theta=0$). Therefore,
under a force $\force$, the position of segments with $\theta>0$
are maintained mostly through entropic mechanisms that prevent their
rotation. 

To determine the response of the chains, the potential energy $\energymon$
of a Kuhn segment must be specified. Subsequently, we can prescribe
(A) the end-to-end distance $\etoe$ or (B) the force $\force$ on
the chain, and compute the corresponding $\beta$ through the following
procedures: 

\subsection{Procedure A\label{subsec:Procedure-A}}

In the case of a prescribed $\etoe$, Eq. \ref{eq:equation_for_bond_length}
allows us to solve for the deformed lengths of the Kuhn segments along
the chains as a function of $\beta$ and $\theta$. Subsequently,
one can plug this relation into Eq. \ref{eq:end-to-end_equation}
to numerically determine the $\beta$. 

\subsection{Procedure B\label{subsec:Procedure-B}}

Alternatively, one can prescribe $\force$, compute $\beta=\force\lenref/\left(\boltzmann T\right)$,
and then employ Eqs. \ref{eq:equation_for_bond_length} and \ref{eq:end-to-end_equation}
to determine the deformed length of the Kuhn segments and the end-to-end
distance $\etoe$ (in this case, $\beta$ is determined as a function
of $\etoe$).

\section{The energy of a repeat unit along the chain}

\begin{figure}[t]
\hspace*{\fill}\includegraphics[width=8cm]{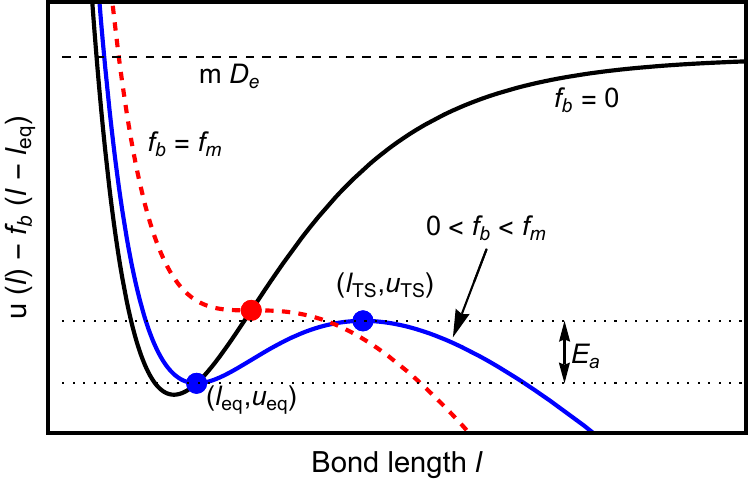}\hspace*{\fill}

\caption{The potential energy landscape $\protect\energymon\left(\protect\lencur\right)-\protect\forcemon\left(\protect\lencur-\protect\leneq\right)$
as a function of the bond length $\protect\lencur$. \label{fig:energy_landscape}}
\end{figure}

Next we explore the enthalpic energy of the chain. Each Kuhn segment
comprises $\bonds$ bonds (typically chemical C-C bonds) with an initial
and a deformed length of $\lenref/\bonds$ and $\lencur/\bonds$,
respectively. In the following, we assume that each bond follows the
tilted Morse-based potential \citep{Beyer2000}, and accordingly the
total potential energy of a Kuhn segment is 
\begin{equation}
\energymon\left(\lencur\right)=\bonds\De\left(1-\exp\left(-\frac{2\forcemax}{\De}\left(\frac{\lencur}{\bonds}-\frac{\lenref}{\bonds}\right)\right)\right)^{2}.\label{eq:morse_potential}
\end{equation}
Here, $\De$ is the dissociation energy, or the energy required to
break a bond in the absence of an external force, and $\forcemax\ge\forcemon$
is the maximum force that can be applied to a bond prior to breaking.
Once an external force is applied (i.e. $\forcemon>0$), bonds stretch
and alter the energy landscape. Through integration of Eq. \ref{eq:force_on_bond},
the potential energy can be written as $\energymon\left(\lencur\right)-\forcemon\left(\lencur-\leneq\right)$,
where $\leneq$ denotes the bond length at equilibrium. This energy
generally has two extrema points - a minimum, corresponding to the
equilibrium state in which the bond length is $\lencur=\leneq$, and
a maximum, which describes the transition state of a bond rupture
event. Fig. \ref{fig:energy_landscape} plots the potential energy
landscape $\energymon\left(\lencur\right)-\forcemon\left(\lencur-\leneq\right)$
as a function of the bond length $\lencur$ in the absence of a force
(black curve), under an applied force $0<\forcemon<\forcemax$ (blue
curve), and for $\forcemon=\forcemax$ (red dashed curve).

The extrema under a force $\forcemon$ can be determined from Eq.
\ref{eq:force_on_bond}, 
\begin{equation}
\lencur=\lenref+\frac{\bonds\De}{2\forcemax}\ln\left(\frac{2\left(1\pm\sqrt{1-\frat}\right)}{\frat}\right),\label{eq:extrema_length}
\end{equation}
where the plus and the minus signs correspond to the transition state
length $\leneqmax$ and the equilibrium length $\leneq$, respectively,
and we define the normalized force $0\le\frat=\forcemon/\forcemax\le1$.
Note that in the limit $\forcemon\rightarrow0$, the potential does
not have a maximum and the equilibrium length $\leneq=\lenref$. It
is convenient to define the energies at the extrema points $\uTS=\energymon\left(\leneqmax\right)-\forcemon\left(\leneqmax-\leneq\right)$
and $\ueq=\energymon\left(\leneq\right)$. 

\begin{figure}
\hspace*{\fill}(a) \includegraphics[width=7cm]{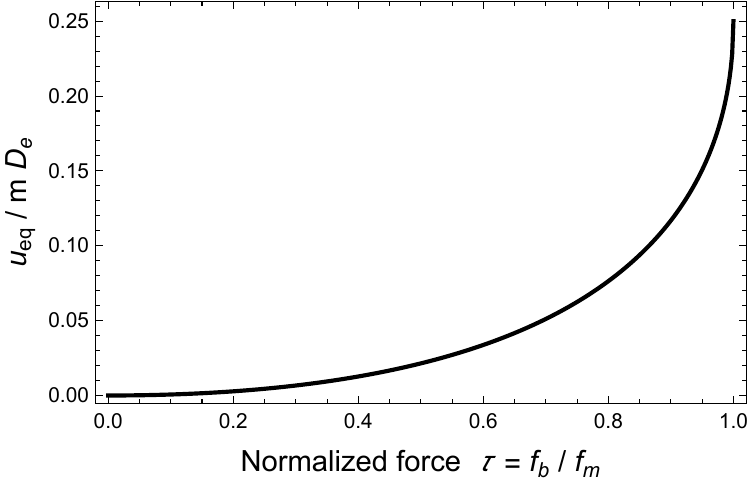}~~~(b)
\includegraphics[width=7cm]{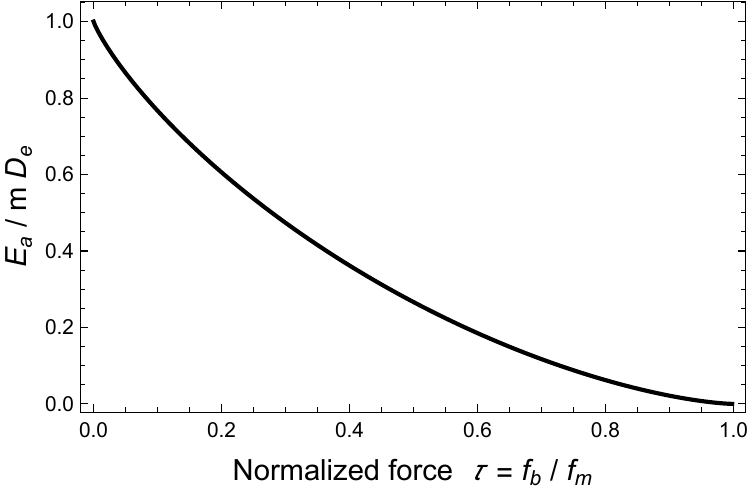}\hspace*{\fill}

\caption{(a) The normalized equilibrium potential $\protect\energymon_{eq}/\protect\bonds\protect\De$
and (b) the normalized activation energy $\protect\Eactivation/\protect\bonds\protect\De$
as a function of the normalized force a bond experiences $\protect\frat=\protect\forcemon/\protect\forcemax$.
\label{fig:supplemental_figures}}
\end{figure}

The rupture of a Kuhn segment depends on the activation energy (or
energy barrier) between the equilibrium and the transition states
\citep{Beyer2000}
\begin{equation}
\Eactivation=\uTS-\ueq=\bonds\De\left(\sqrt{1-\frat}+\frac{\frat}{2}\ln\left(\frac{1-\sqrt{1-\frat}}{1+\sqrt{1-\frat}}\right)\right),\label{eq:activation_energy}
\end{equation}
where Eq. \ref{eq:extrema_length} is used. Specifically, for a given
normalized force $\frat$, larger $\Eactivation$ corresponds to a
low probability of a rupture, and once the maximum force is applied
(i.e. $\forcemon=\forcemax$), $\leneqmax=\leneq$ the bond breaks
spontaneously. Note that the activation energy $\Eactivation$ decreases
monotonically, with the limiting values $\Eactivation=\bonds\De$
and $\Eactivation=0$ for $\forcemon\rightarrow0$ and $\forcemon=\forcemax$,
respectively. Figs. \ref{fig:supplemental_figures}a and \ref{fig:supplemental_figures}b
plot the normalized equilibrium potential $\energymon_{eq}/\bonds\De$
and the activation energy $\Eactivation/\bonds\De$ as a function
of $\frat$, respectively. 

Chain scission occurs when one of the bonds in a Kuhn segment along
the chain breaks. To capture the stochastic nature of chain failure,
we define the probability to break a single bond along a Kuhn segment
subjected to a normalized force $\frat$ by 

\begin{equation}
\pbond\left(\frat\right)=\exp\left(-\frac{\Eactivation\left(\frat\right)}{\bonds\,\boltzmann T}\right).\label{eq:probability_bond}
\end{equation}
Note that $0<\pbond\le1$, with the limits $\pbond\left(\Eactivation\left(\frat\right)\rightarrow\infty\right)\rightarrow0$
and $\pbond\left(\Eactivation\left(\frat\right)=0\right)=1$. Therefore,
the probability of chain rupture is
\begin{equation}
\pchain\left(\force\right)=1-p_{NR},\label{eq:pc_general}
\end{equation}
where 
\begin{align}
p_{NR} & =\Pi_{i}\left(1-\pbond^{\left(i\right)}\left(\frat^{\left(i\right)}\right)\right)\Rightarrow\ln p_{NR}=\sum_{i}\ln\left(1-\pbond^{\left(i\right)}\left(\frat^{\left(i\right)}\right)\right)\nonumber \\
 & \Rightarrow p_{NR}=\exp\left(\monomers\int_{0}^{\pi}\ln\left(1-\pbond\left(\frat\right)\right)\pdf\solidangle\right),\label{eq:p_NR}
\end{align}
is the probability that none of the bonds break. In Eq. \ref{eq:p_NR}
the products and summations are carried over all Kuhn segments and
accordingly $p_{NR}$ directly depends on the number of Kuhn segments
in the chain. Specifically, for a constant maximum force $\forcemax$
longer chains have a higher probability of breaking at lower forces
$\force$ since there are more segments that can dissociate. In addition,
we account for the distribution of forces along the chain, i.e. each
Kuhn segment experiences the normalized force $\frat$. It is once
again emphasized that the force experienced by the Kuhn segments depends
on their orientation with respect to the end-to-end distance (recall
$\frat\sim\cos\theta$ through $\forcemon$). Substitution of Eq.
\ref{eq:p_NR} into Eq. \ref{eq:pc_general} yields 
\begin{equation}
\pchain\left(\force\right)=1-\exp\left(\monomers\int_{0}^{\pi}\ln\left(1-\pbond\left(\frat\right)\right)\pdf\solidangle\right),\label{eq:probability_rupture_chain}
\end{equation}
where $\pbond$ is given by Eq. \ref{eq:probability_bond}. 

Before proceeding, we point out that the proposed formulation is independent
of the rate of loading. While it is possible to include a time scale
to account for the kinetics, in general the time scale of breaking
bonds is small in comparison with the loading time scale. Therefore,
we argue that a rate independent approach is a simple and good approximation
for chain scission, especially at low strain rates.

For convenience, we summarize all of the model parameters in Table
\ref{tab:summary_of_model_parameters}.

\begin{table}
\caption{Summary of model parameters. \label{tab:summary_of_model_parameters}}

\hspace*{\fill}%
\begin{tabular}{cc}
\toprule 
Notation & Significance\tabularnewline
\midrule
\midrule 
$\etoe$ & End-to-end distance\tabularnewline
\midrule 
$\monomers$ & Number of Kuhn segments in chain\tabularnewline
\midrule 
$\lenref,\lencur$ & initial, deformed length of Kuhn segment\tabularnewline
\midrule 
$\energymon$ & Potential energy of Kuhn segment\tabularnewline
\midrule 
$\rat=\etoe/\monomers\lenref$ & Ratio between $\etoe$ and initial contour length of chain\tabularnewline
\midrule 
$\force,\forcemon$ & Force on chain, Kuhn segment\tabularnewline
\midrule 
$\forcemax$ & Maximum force on Kuhn segment\tabularnewline
\midrule 
$\frat=\forcemon/\forcemax$ & Ratio between force on Kuhn segment and maximum force\tabularnewline
\midrule 
$\beta$ & Lagrange multiplier enforcing Eq. \ref{eq:constraint2}, determines
the force on a chain\tabularnewline
\midrule 
$\De$ & Dissociation energy required to break bond\tabularnewline
\midrule 
$\bonds$ & Number of bonds in a Kuhn segment\tabularnewline
\midrule 
$\Eactivation$ & Activation energy (energy barrier)\tabularnewline
\midrule 
$\pbond,\pchain$ & probability of breaking a bond, chain\tabularnewline
\midrule 
$\stressT,\piolaT$ & True (Cauchy), first Piola-Kirchhoff stress tensor\tabularnewline
\bottomrule
\end{tabular}\hspace*{\fill}
\end{table}

\section{Mechanical response of single chains}

\subsection{The classical chain}

To demonstrate the merit of the proposed framework, we consider a
chain characterized by the maximum force $\forcemax=5\,\mathrm{nN}$
with $\monomers=40$ Kuhn segments of length $\lenref=1\,\mathrm{nm}$,
where each Kuhn segment has $\bonds=5$ chemical C-C bonds with $\De=150\,\boltzmann T$
\citep{Wang2002,Beyer2000,Zhang2003,Wang2019}. The chain is stretched
such that its end-to-end distance $\etoe$ increases from $0$ to
rupture. 

\begin{figure}
\hspace*{\fill}(a) \includegraphics[width=7cm]{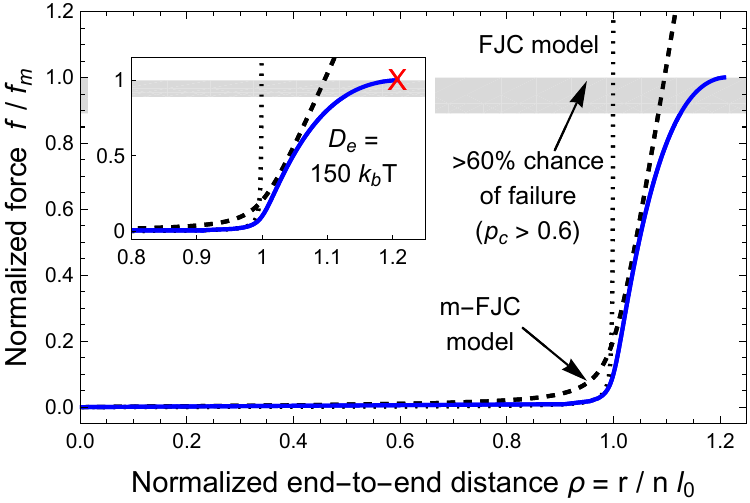}~~~(b)
\includegraphics[width=7cm]{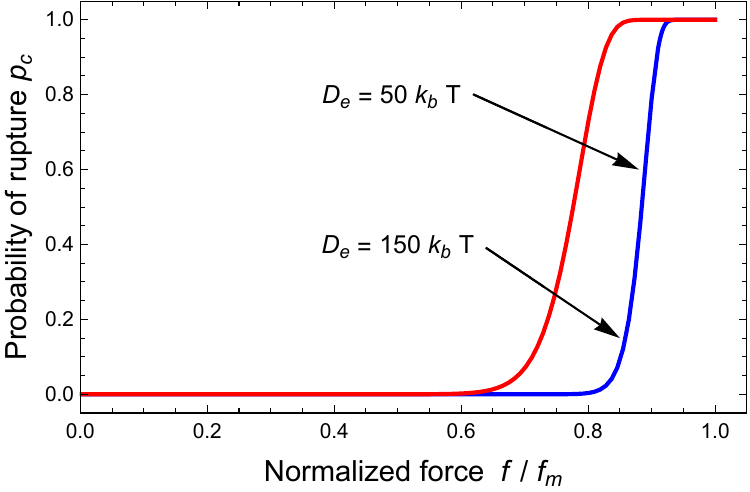}\hspace*{\fill}

\caption{(a) The normalized force on the chain $\protect\force/\protect\forcemax$
as a function of the normalized end-to-end distance $\protect\rat=\protect\etoe/\protect\monomers\protect\lenref$.
(b) The probability of rupture $\protect\pchain$ as a function of
the normalized chain force $\protect\force/\protect\forcemax$. \label{fig:chain_behavior}}
\end{figure}

Fig. \ref{fig:chain_behavior}a plots the normalized force on the
chain $\force/\forcemax$ as a function of the normalized end-to-end
distance $\rat=\etoe\,/\,\monomers\,\lenref$. The force $\force$
is calculated via Eq. \ref{eq:force_chain} through procedure A (Sec.
\ref{subsec:Procedure-A}) using the tilted Morse potential (Eq. \ref{eq:morse_potential}).
It is worth noting that Kuhn segments along the chain experience a
force $\forcemon\le\force$, with $\forcemon=\force$ for segments
that are aligned along the end-to-end vector (i.e $\theta=0$). The
dotted and the dashed curves denote the predictions according to the
FJC model with rigid Kuhn segments and the modified FJC (m-FJC), in
which $\etoe/\monomers\lenref=\left(\coth\left(\force/\force_{c}\right)-\force_{c}/\force\right)\left(1+\force/\force_{s}\right)$
\citep{Smith1996,Zhang2003}. The latter accounts for the entropic
and enthalpic responses at low and high forces, with $\force_{c}\approx\boltzmann T/\lenref$
and $\force_{s}$ as the characteristic conformational and stretching
tensions, respectively. In Fig. \ref{fig:chain_behavior}a, we follow
experimental findings and set $\force_{c}\approx4\,\mathrm{pN}$ and
$\force_{s}\approx10\,\mathrm{nN}$ and show that the present model
follows the trend of the m-FJC. It is worth mentioning that the m-FJC
overestimates the chain force and does not account for the limiting
rupture force. The introduction of the tilted Morse potential in the
proposed model includes a maximum force and therefore provides a practical
physically-motivated tool to predict chain scission. The shaded region
marks the forces for which the probability of failure is $>60\%$. 

To better understand the stochastic nature of chain scission, we examine
the probability of chain scission for strong chemical bonds and weaker
physical bonds. Typically, physical bonds are characterized by low
dissociation energies ($\De\sim1-50\,\boltzmann T$) and a maximum
force of $\forcemax\sim10-100\,\mathrm{pN}$ \citep{Sheu2003,Du2011,Beyer2000}.
Fig. \ref{fig:chain_behavior}b plots the probability of rupture $\pchain$
(Eq. \ref{eq:probability_rupture_chain}) as a function of the normalized
chain force $\force/\forcemax$ for strong chemical ($\De=150\,\boltzmann T$)
and weak physical ($\De=50\,\boltzmann T$) bonds. We find that the
probability of failure for chains comprising strong chemical bonds
with a high dissociation energy $\De$ is very small for $\force/\forcemax<0.9$,
and increases as forces near $\forcemax$. Weaker bonds have a higher
probability of breaking at forces that are smaller relative to their
corresponding maximum force $\forcemax$. This is particularly relevant
for systems with sacrificial physical bonds, hidden length-type mechanisms,
and mechanophore activation \citep{Ducrot2014,Zhang2023,Du2011,Olive2024}.
Note that rupture is guaranteed ($\pchain=1$) once the maximum force
$\force=\forcemax$ is achieved.

\subsection{A polymer chain with sacrificial bonds}

\begin{figure}
\hspace*{\fill}(a) \includegraphics[width=10cm]{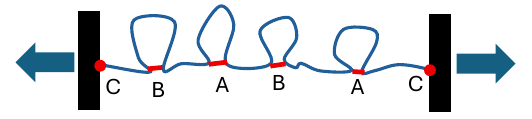}\hspace*{\fill}

\hspace*{\fill}(b) \includegraphics[width=10cm]{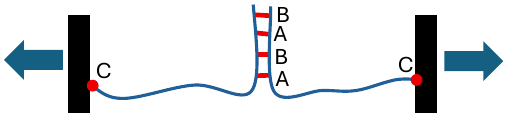}\hspace*{\fill}

\caption{Several sacrificial bonds on single chains that are connected to substrates
via bond C. (a) A molecule with four fixed internal loops through
the sacrificial bonds A and B. Once a sufficient force is applied,
a sacrificial bond breaks to reveal the corresponding hidden length
from the free loop. (b) Two chain segments that are interconnected
through a sequence of four sacrificial bonds (of types A and B). The
application of a sufficiently large force gradually breaks the bonds
to extend the length of the chain. Once all bonds dissociate, the
chains break. \label{fig:SB_systems}}
\end{figure}

\begin{figure}
\hspace*{\fill}\includegraphics[width=8cm]{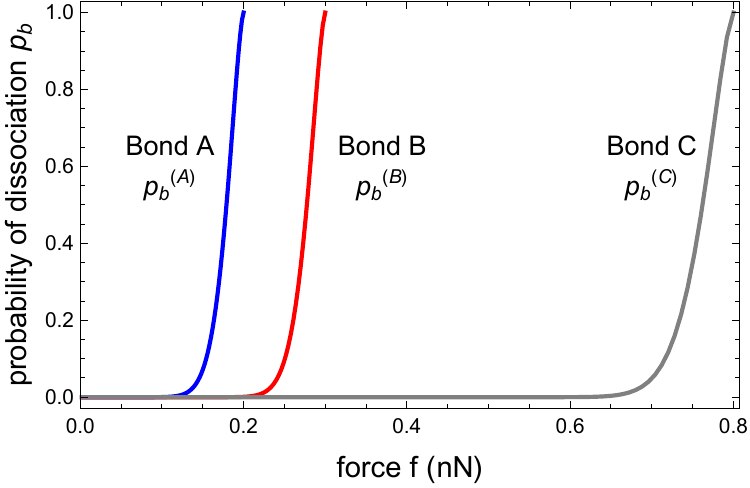}\hspace*{\fill}

\caption{The probability of bond dissociation $\protect\pbond^{\left(\bullet\right)}$
as a function of the force $\protect\force$ for the bonds $\bullet=A$,
$\bullet=B$, and $\bullet=C$. \label{fig:SB_probability}}
\end{figure}

Chains with sacrificial bonds are common in nature and synthetic systems
\citep{Du2011,Cohen2019a,Cohen2021a,Zhang2003,Chen&etal05macromol,Olive2024}.
Typically, these bonds can be ionic bonds, hydrogen bonds, hydrophobic
interactions, or Van der Waals \citep{Fantner2006,Cohen2026,Olive2025}.
In the following, we employ our framework to investigate the influence
of sacrificial bonds on two types of chain configurations with three
sacrificial bonds, illustrated in Fig. \ref{fig:SB_systems}. Here,
bonds A, B, and C are characterized by the dissociation energies $\De^{\left(A\right)}=30\,\boltzmann T$,
$\De^{\left(B\right)}=50\,\boltzmann T$, and $\De^{\left(C\right)}=100\,\boltzmann T$,
and the maximum forces $\forcemax^{\left(A\right)}=0.21\,\mathrm{nN}$,
$\forcemax^{\left(B\right)}=0.31\,\mathrm{nN}$, $\forcemax^{\left(C\right)}=0.82\,\mathrm{nN}$.
The segments between the sacrificial bonds are assumed to be much
stronger and therefore are not likely to break. For convenience, Fig.
\ref{fig:SB_probability} depicts the probability of dissociation
of the sacrificial bonds $A$, $B$, and $C$, as given by Eq. \ref{eq:probability_bond}
with $\bonds=1$. 

Fig. \ref{fig:SB_systems}a describes a chain with internal loops
that reveal a hidden length once the sacrificial bonds break. As expected,
the bonds break in sequential order, based on their corresponding
maximum force $\forcemax$. Note that once bonds A and B break, failure
can occur through chain scission or the breaking of the interactions
between the chain and the stretching device (bond C). Such configurations
are common in single protein-based chain experiments with atomic force
microscopy \citep{Li2000}. Since we assume that the intramolecular
interactions are much stronger, detachment through the dissociation
of bond C is expected to occur. Fig. \ref{fig:SB_systems}b illustrates
a chain with two parts that are interconnected by parallel bonds.
Once sufficient force is applied, the bonds break successively according
to the order at which they are arranged. This configuration is common
in the unzipping of DNA \citep{Krautbauer2003,Bockelmann2002}, where
the hidden length that is released per bond dissociation is the distance
between binding sites. 

To demonstrate the behavior of these two systems, we consider the
chains that are initially characterized by $\monomers=45$ effective
repeat units of length $\lenref=1\,\mathrm{nm}$. Each bond hides
a loop segment comprising an additional $45$ repeat units. Therefore,
the dissociation of a bond adds $45$ repeat units to the overall
contour length such that the overall contour length is $\monomers=45+45x$,
where $x$ is the number of broken bonds. 

The two systems are subjected to uniaxial extension under two types
of boundary conditions: (1) displacement controlled loading and (2)
force controlled loading. To determine the relation between the end-to-end
distance and the applied force, we assume that the bonds experience
the macroscopically applied force and calculate the probability of
rupture $\pbond^{\left(\bullet\right)}$ through Eq., where $\bullet=A,B,$
or $C$. Next, a random number $0\le\nu\le1$ is generated and dissociation
occurs if $\pbond^{\left(\bullet\right)}>\nu$.

\begin{figure}
\hspace*{\fill}(a) \includegraphics[width=7cm]{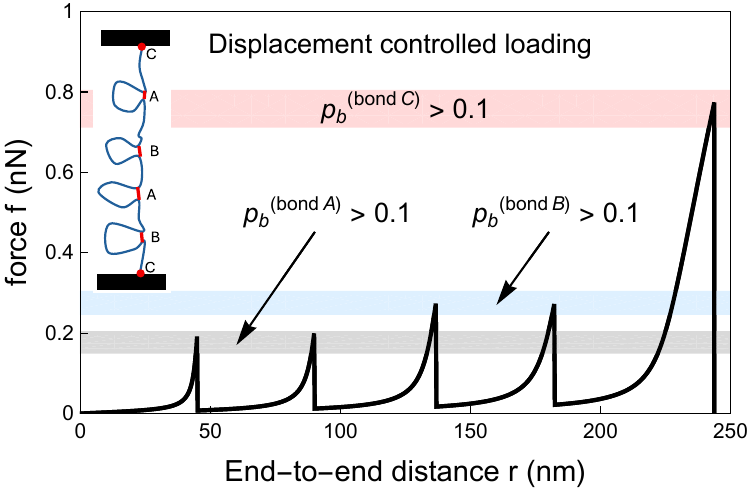}~~~(b)
\includegraphics[width=7cm]{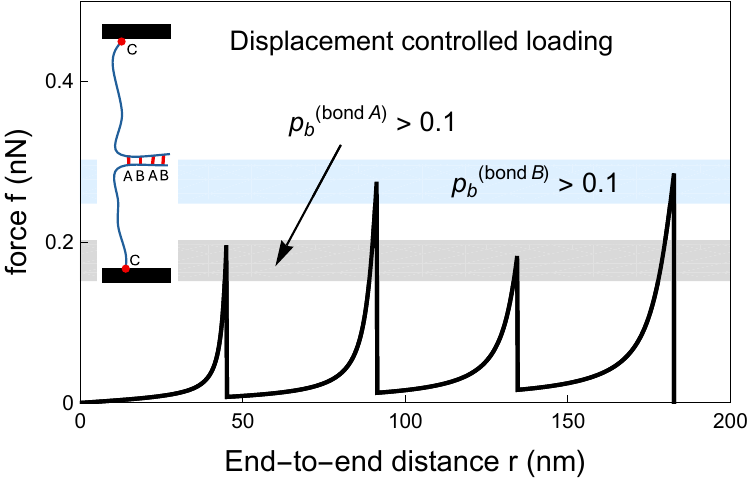}\hspace*{\fill}

\caption{The force $\protect\force$ as a function of the end-to-end distance
$\protect\etoe$ under displacement controlled loading for systems:
(a) a molecule with four fixed internal loops through the sacrificial
bonds A and B and (b) two chain segments that are interconnected through
a sequence of four sacrificial bonds (of types A and B). The shaded
regions denote a probability of bond dissociation $\protect\pbond>0.1$.
\label{fig:SB_disp_controlled}}
\end{figure}

Figs. \ref{fig:SB_disp_controlled}a and \ref{fig:SB_disp_controlled}b
depict the external force $\force$ as a function of the end-to-end
distance $\etoe$ to demonstrate the response of the two systems under
displacement controlled setting (following procedure A, Sec. \ref{subsec:Procedure-A}).
The shaded regions mark a force for which $\pbond^{\left(\bullet\right)}>0.1$,
or $>10\%$ bond rupture, where $\bullet=A,B,$ or $C$. We point
out that the relaxation of the force on the chain after a sacrificial
bond breaks is proportional to the number of segments (or degrees
of freedom) that are added due to the hidden length. To understand
this, note that the added segments provide additional mobility that
increases the entropy, and therefore decrease the applied force. 

\begin{figure}
\hspace*{\fill}(a) \includegraphics[width=7cm]{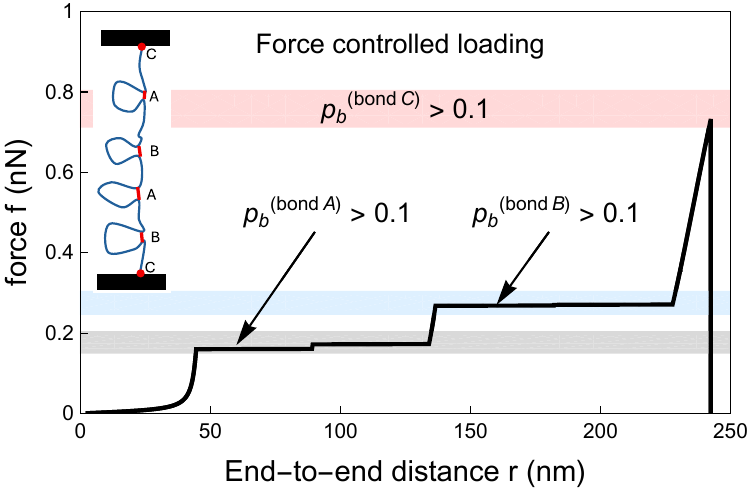}~~~(b)
\includegraphics[width=7cm]{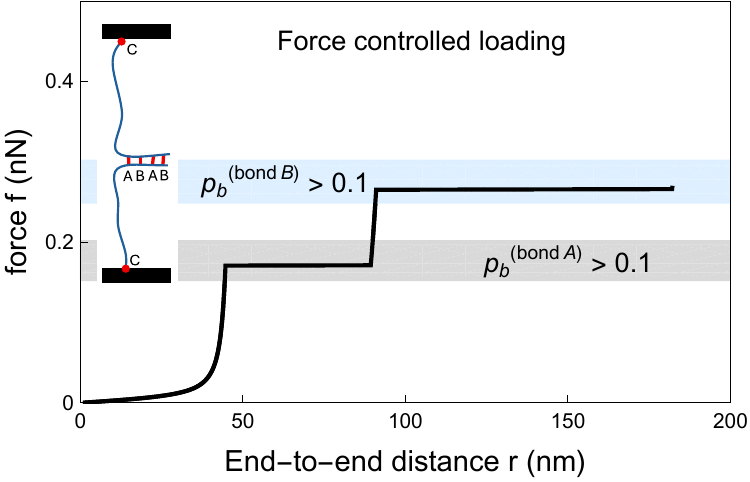}\hspace*{\fill}

\caption{The force $\protect\force$ as a function of the end-to-end distance
$\protect\etoe$ under force controlled loading for systems: (a) a
molecule with four fixed internal loops through the sacrificial bonds
A and B and (b) two chain segments that are interconnected through
a sequence of four sacrificial bonds (of types A and B). The shaded
regions denote a probability of bond dissociation $\protect\pbond>0.1$.
\label{fig:SB_force_controlled}}
\end{figure}

Next, we plot the force $\force$ as a function of the end-to-end
distance $\etoe$ of the two systems under force controlled setting
in Figs. \ref{fig:SB_force_controlled}a and \ref{fig:SB_force_controlled}b.
Here, we follow procedure B described in Sec. \ref{subsec:Procedure-B}.
In this case, the two chains behave differently. In the case of a
molecule with four fixed internal loops, once a critical force $\sim\forcemax^{\left(A\right)}$
is applied, the two bonds of type A break (almost simultaneously)
and a sudden ``snap-through'', or an immediate extension of the
chain, occurs. This effect is the result of the boundary conditions
- since the force is not allowed to relax, the added repeat units
from the hidden length are not allowed to relax, the entropy cannot
increase, and accordingly the chain must extend abruptly to accommodate
the external loading. A similar effect occurs with the two bonds of
type B near the force $\sim\forcemax^{\left(B\right)}$. Further increase
in loading leads to additional extension up to the force $\forcemax^{\left(C\right)}$,
at which point the chain detaches.

Examination of the two chain segments that are interconnected through
a sequence of four sacrificial bonds (as shown in Fig. \ref{fig:SB_systems}b)
reveals a different trend. The external force ramps up to $\sim\forcemax^{\left(A\right)}$,
at which point the first bond of type A dissociates and the chain
instantaneously stretches. A further increase in force is required
to dissociate the next bond in the sequence (bond of type B). Once
the force $\sim\forcemax^{\left(B\right)}$ is reached, this bond
breaks. Since a monotonically increasing force is prescribed and $\forcemax^{\left(B\right)}>\forcemax^{\left(A\right)}$,
the next two bonds immediately break as well and the chain loses its
structural integrity. 

The above examples emphasize the crucial role of boundary conditions
in systems with sacrificial bonds - prescribing displacement allows
the system to relax and regain its chain-like properties, whereas
monotonic manipulation of the force leads to snap-through-like deformations
and abrupt scission / detachment. 

\global\long\def\refchains{N_{0}}%
\global\long\def\refchainss{N_{0}^{\left(s\right)}}%
\global\long\def\refchainsc{N_{0}^{\left(c\right)}}%
\global\long\def\refchainsl{N_{0}^{\left(l\right)}}%

\global\long\def\chains{N}%
\global\long\def\chainss{N^{\left(s\right)}}%
\global\long\def\chainsc{N^{\left(c\right)}}%
\global\long\def\chainsl{N^{\left(l\right)}}%

\global\long\def\mons{\monomers^{\left(s\right)}}%
\global\long\def\monc{\monomers^{\left(c\right)}}%
\global\long\def\monl{\monomers^{\left(l\right)}}%

\global\long\def\lenrefs{L^{\left(s\right)}}%
\global\long\def\lenrefc{L^{\left(c\right)}}%
\global\long\def\lenrefl{L^{\left(l\right)}}%

\global\long\def\disp{\delta}%
\global\long\def\stretch{\lambda}%

\section{Implementation to 3-dimensional networks}

\global\long\def\stretchch{\lambda_{c}}%
\global\long\def\inone{I_{1}}%
\global\long\def\freech{\psi_{c}}%
\global\long\def\free{\psi}%
\global\long\def\pressure{p_{r}}%
\global\long\def\damage{d}%
\global\long\def\locdamage{\tilde{d}}%

\global\long\def\RT{\mathbf{R}}%
\global\long\def\rT{\mathbf{r}}%
\global\long\def\RTdir{\hat{\mathbf{R}}}%
\global\long\def\rTdir{\hat{\mathbf{r}}}%
\global\long\def\defgradT{\mathbf{F}}%

\global\long\def\xh{\mathbf{\hat{x}}}%
\global\long\def\yh{\mathbf{\hat{y}}}%
\global\long\def\zh{\mathbf{\hat{z}}}%
\global\long\def\angch{\theta_{ch}}%
 
\global\long\def\phich{\phi_{ch}}%

\global\long\def\stressC{\boldsymbol{\sigma}_{c}}%
\global\long\def\stress{\sigma}%
\global\long\def\Piola{P}%
\global\long\def\solidanglech{\solidangle_{ch}}%

The proposed local chain model can be integrated into a 3-dimensional
network to capture the macroscopic response of polymer networks and
gels subjected to general loading conditions. In the following, we
demonstrate this in the context of (1) a 3-dimentional network with
an isotropic distribution of chains and (2) the eight-chain model
of \citet{Arruda1993}. To capture the effect of chain scission in
the network, we define a coordinate system $\left\{ \xh,\yh,\zh\right\} $
such that the referential direction of a chain can be written as
\begin{equation}
\RTdir\left(\angch,\phich\right)=\cos\angch\xh+\sin\angch\left(\cos\phich\yh+\sin\phich\zh\right),
\end{equation}
where $0\le\angch\le\pi$ and $0\le\phich<2\pi$ are the polar and
the azimuthal angles, respectively. Next, we define the damage parameter
\begin{equation}
\damage=\int_{0}^{2\pi}\int_{0}^{\pi}\pchain\,\solidanglech,\label{eq:damage_general}
\end{equation}
where $\solidanglech=\left(\sin\angch/4\pi\right)\d\angch\d\phich$
is the normalized solid angle which accounts for the number of chains
along the direction $\RTdir\left(\angch,\phich\right)$ and $\pchain\left(\force\right)$
is the probability of breaking a chain along the direction $\RTdir\left(\angch,\phich\right)$
that is subjected to a force $\force=\force\left(\defgradT,\angch,\phich\right)$,
given in Eq. \ref{eq:probability_rupture_chain}. We emphasize that
the force $\force$ depends on the orientation of the chain through
the deformation gradient. In addition, while in the following we focus
our attention to uniaxial loading, the damage parameter $\damage$
describes the number of chains that dissociated and therefore it can
only increase. This is particularly important during cyclic loading,
in which the damage $\damage$ remains constant during the unloading
phase. 

The model parameters used in the following are $\monomers=10$, $\lenref=0.6\,\mathrm{nm}$,
$\bonds=5$, $\forcemax=7\,\mathrm{nN}$, and $\De=145\,\boltzmann T$. 

\subsection{An isotropic network of chains}

We begin by considering a network with randomly oriented and uniformly
distributed chains. As before, each chain comprises $\monomers$ repeat
units with an initial length $\lenref$. The chain-density along the
direction $\RTdir^{\left(i\right)}$ is $\refchains^{\left(i\right)}=\refchains\solidanglech$,
and the referential end-to-end vector of the $i$-th chain is $\RT^{\left(i\right)}=R\RTdir^{\left(i\right)}$,
where $R=\sqrt{\monomers}\lenref$.

The network is subjected to a macroscopic deformation characterized
by the deformation gradient $\defgradT$. We assume that the chains
experience affine deformations such that the deformed end-to-end vector
of the $i$-th chain is $\rT^{\left(i\right)}=\defgradT\RT^{\left(i\right)}$.
The end-to-end distance and the direction of this chain are $r^{\left(i\right)}=\sqrt{\rT^{\left(i\right)}\cdot\rT^{\left(i\right)}}$
and $\rTdir^{\left(i\right)}=\rT^{\left(i\right)}/r^{\left(i\right)}$.
In the deformed state, the chain-density along the $i$-th direction
is $\chains^{\left(i\right)}=\left(1-\pchain^{\left(i\right)}\right)\refchains^{\left(i\right)}$,
where $\pchain^{\left(i\right)}=\pchain\left(\angch^{\left(i\right)},\phich^{\left(i\right)},\defgradT\right)$
depends on the orientation of the chain and the deformation gradient. 

The energy of the network is 
\begin{equation}
\free=\refchains\int_{0}^{2\pi}\int_{0}^{\pi}\left(1-\pchain\right)\freech\solidanglech,
\end{equation}
and the total true stress is
\begin{equation}
\stressT=\refchains\int_{0}^{2\pi}\int_{0}^{\pi}\left(1-\pchain\right)\stressC\solidanglech-\pressure\mathbf{I},
\end{equation}
where we define the stress associated with the $i$-th chain 
\begin{equation}
\stressC^{\left(i\right)}=\frac{\partial\freech}{\partial\defgradT}\defgradT^{T}=\boltzmann T\,\monomers\,\beta\left(\rat^{\left(i\right)}\right)\rat^{\left(i\right)}\rTdir\otimes\rTdir.
\end{equation}
Here, we recall that $\rat^{\left(i\right)}=\etoe/\monomers\lenref$
is the ratio between the end-to-end distance and the contour length
and $\pressure$ is the work-less pressure-like term that is determined
from the boundary conditions and enforces the incompressibility of
the network. The first Piola-Kirchhoff stress is $\piolaT=\stressT\defgradT^{-T}$.
We also compute the total number of chains
\begin{equation}
\chains=\refchains\int_{0}^{2\pi}\int_{0}^{\pi}\left(1-\pchain\right)\solidanglech,
\end{equation}
which can related to the damage via $\damage=1-\chains/\refchains$
through Eq. \ref{eq:damage_general}. Here, $\refchains$ and $\chains$
are the number of chains per unit referential volume. 

To demonstrate the response and damage that accumulates in a network,
we focus on incompressible polymers and employ the formulations to
the special case of uniaxial extension, characterized by the deformation
gradient 
\begin{equation}
\defgradT_{u}=\stretch\xh\otimes\xh+\frac{1}{\sqrt{\stretch}}\left(\yh\otimes\yh+\zh\otimes\zh\right).\label{eq:defgrad_uniaxial}
\end{equation}
Note that under uniaxial deformation, one can rewrite $\pchain\left(\force\right)=\pchain\left(\stretch,\angch,\phich\right)$.
The boundary conditions are prescribed through $\stretch$ and the
traction free boundaries $\stressT\yh\cdot\yh=\stressT\zh\cdot\zh=0$,
which enable to determine the pressure term $\pressure$. To perform
the integration from the chain to the network level, we employ the
micro-sphere technique described in Appendix \ref{sec:microsphere}
with $42$ representative directions (given in Table 1 of \citet{Bazant&oh86ZAMM}). 

\begin{figure}
\hspace*{\fill}(a) \includegraphics[width=6.5cm]{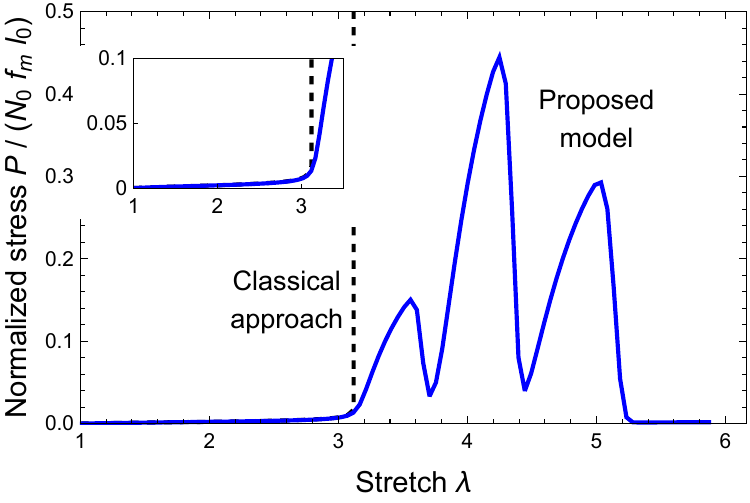}\hspace*{\fill}

\hspace*{\fill}(b) \includegraphics[width=6.5cm]{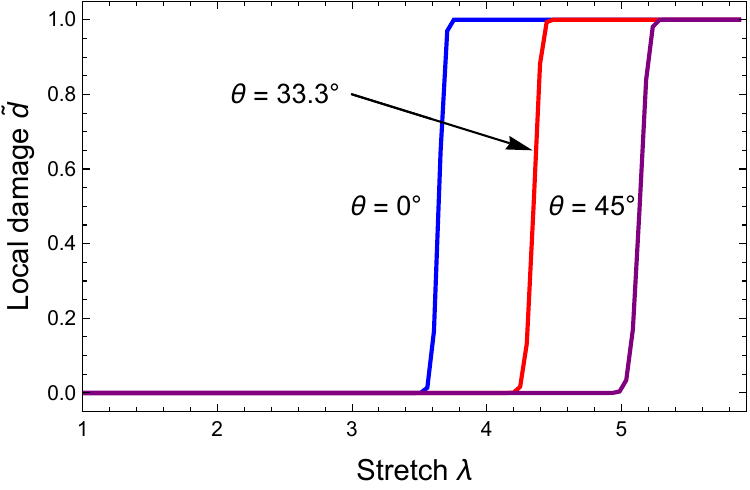}~~~(c)
\includegraphics[width=6.5cm]{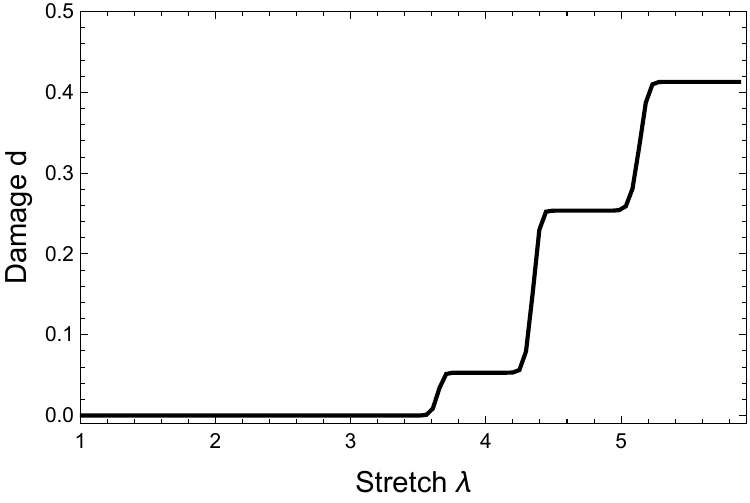}\hspace*{\fill}

\caption{(a) The normalized stress $\protect\Piola/\left(\protect\refchains\protect\forcemax\protect\lenref\right)$,
(b) the local damage $\protect\locdamage$, and (c) the global damage
$\protect\damage$ as a function of the uniaxial stretch $\protect\stretch$.
\label{fig:micro-sphere}}
\end{figure}

Fig. \ref{fig:micro-sphere}a plot the normalized stress $\Piola/\left(\refchains\forcemax\lenref\right)$
as a function of the stretch $\stretch$. Since the chains are characterized
by a limiting force, the normalization $\refchains\forcemax\lenref=\beta_{m}\refchains\boltzmann T$,
where $\beta_{m}=\forcemax\lenref/\boltzmann T$ denotes the maximum
value of $\beta$ at the rupture force $\forcemax$, is used for convenient.
The dashed curve corresponds to the response according to the classical
theory, where $\beta$ is determined from the Langevin function, and
the continuous curve is the proposed model. 

To demonstrate the progressive damage in the network, we note that
the response of chains is symmetric with respect to the stretching
direction $\xh$ and therefore define the local damage of chains characterized
by an angle $\angch$ via $\locdamage\left(\stretch,\angch\right)=\int_{0}^{2\pi}\pchain\left(\stretch,\angch,\phich\right)/2\pi\,\d\phich$.
Fig. \ref{fig:micro-sphere}b plots the local damage $\locdamage$
of a family of chains that form the an angle $\angch$ with respect
to the stretching direction $\xh$. We select the three representative
angles $\angch=0^{\circ},33.3^{\circ},$and $45^{\circ}$, which characterize
the representative directions in the chosen micro-sphere integration
scheme. The total damage $\damage$, as defined in Eq. \ref{eq:damage_general},
is depicted in \ref{fig:micro-sphere}c. 

As expected, the proposed model exhibits the same response as the
classical approaches under small-to-moderate deformations. However,
as the chains stretch close to their contour length, the stress approaches
its limit and rupture begins. First, chain scission occurs in chains
that are aligned along the stretching direction. As further stretch
is applied, more chains rotate towards the $\xh$-direction and elongate,
leading to the ultimate failure of the network. It is worth noting
that the gradual scission events are characterized by a local drop
in stress (or the ``humps'') in Fig. \ref{fig:micro-sphere}a and
as ``steps'' in the damage curve shown in \ref{fig:micro-sphere}c.
However, it is underscored that the micro-sphere-based integration
employed in this work employs $42$ representative directions, characterized
by $5$ angles $\angch$ (see Table 1 in \citet{Bazant&oh86ZAMM}).
As shown by \citet{Mulderrig2021}, consideration of additional chain
orientations through higher quadrature schemes is expected to lead
to a smoother transition in the stress-stretch and the damage-stretch
curves at additional computational cost. 

\subsection{The eight-chain model}

The previous framework, which accounts for the different chain orientations
in the network, is advantageous since it can capture the gradual rupture
of chains and provides a more robust realization of the network. However,
its main drawback is the high computational cost required to obtain
a solution, which increases dramatically as more chain orientations
are considered \citep{Mulderrig2021}. To overcome this, we also propose
to embed the local chain response into the eight-chain topology proposed
by \citet{Arruda1993}. Broadly, this model assumes that the response
of an incompressible network with $\refchains$ referential chains
per unit volume, where each chain comprises $\monomers$ repeat units
of an initial length $\lenref$, can be determined from eight representative
chains that describe the underlying structure. 

The network is subjected to a deformation gradient $\defgradT$ and,
in a deformed state, is has $\chains=\left(1-\damage\right)\refchains$
(the damage parameter $\damage$ is given in Eq. \ref{eq:damage_general}).
The stretch of the eight chains is $\stretchch=\etoe/R=\sqrt{\inone/3}$,
where $\inone=\mathrm{Tr}\left(\defgradT^{T}\defgradT\right)$ is
the first invariant. Subsequently, the free energy of a chain $\freech=\energyC-T\entropyC$
can be computed and the total strain energy-density is $\free=\chains\freech$.
The true stress tensor is then
\begin{equation}
\stressT=\chains\boltzmann T\beta\sqrt{\frac{\monomers}{3\inone}}\defgradT\defgradT^{T}-\pressure\mathbf{I},\label{eq:stress_general_eight_chain}
\end{equation}
where, as before, $\pressure$ is the pressure term that enforces
incompressibility and is determined from the boundary conditions.
Recall that $\beta$ is determined from Eq. \ref{eq:end-to-end_equation}.
Once again, we recall that the first Piola-Kirchhoff stress tensor
$\piolaT=\stressT\defgradT^{-T}$. 

It is emphasized that since the response of the network is based on
the assumption that the eight representative chains experience the
same deformation $\stretch_{c}=\sqrt{\inone/3}$ \citep{Arruda1993},
the damage in Eq. \ref{eq:damage_general} simplifies to $\damage=\pchain$. 

\begin{figure}
\hspace*{\fill}(a) \includegraphics[width=6.5cm]{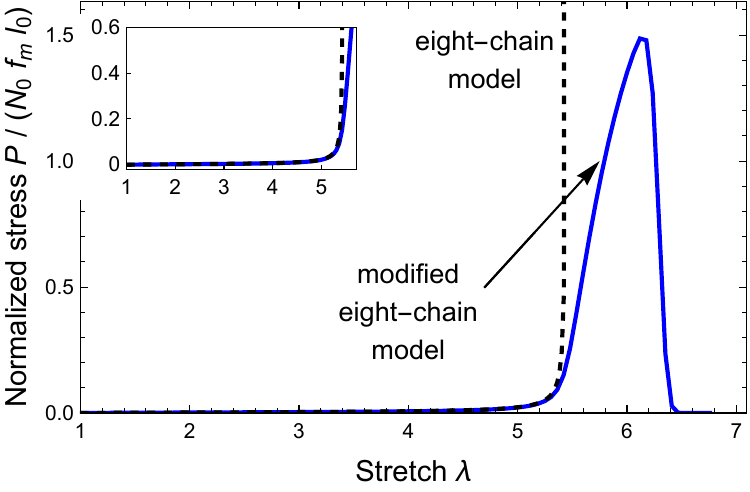}~~~(b)
\includegraphics[width=6.5cm]{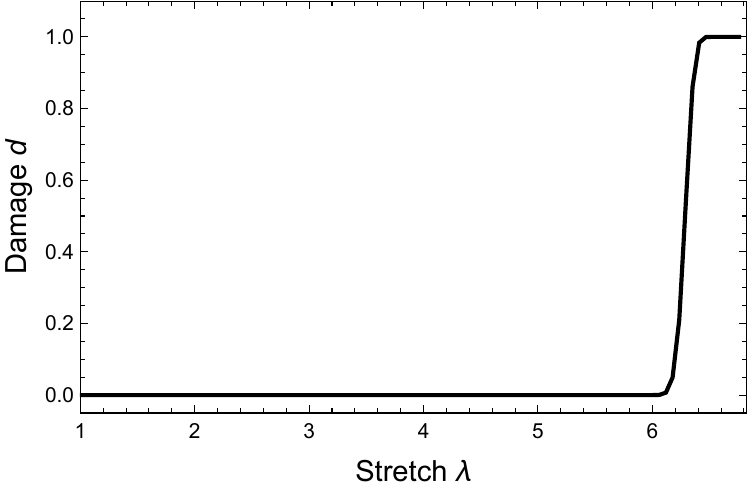}\hspace*{\fill}

\caption{(a) The normalized stress $\protect\Piola/\left(\protect\refchains\protect\forcemax\protect\lenref\right)$
and (b) the damage $\protect\damage$ as a function of the uniaxial
stretch $\protect\stretch$ for a polymer subjected to uniaxial stretch.
\label{fig:eight_chain_model}}
\end{figure}

To demonstrate the response and the damage in the network, we consider
a polymer subjected to uniaxial extension characterized by the deformation
gradient $\defgradT_{u}$ (Eq. \ref{eq:defgrad_uniaxial}). By employing
the boundary conditions $\stressT\yh\cdot\yh=\stressT\zh\cdot\zh=0$,
one can determine the pressure term and the uniaxial nominal stress
\begin{equation}
\Piola=\chains\boltzmann T\beta\sqrt{\frac{\monomers}{3\inone}}\left(\stretch-\frac{1}{\stretch^{2}}\right).\label{eq:eight_chain_uniaxial}
\end{equation}

Figs. \ref{fig:eight_chain_model}a and \ref{fig:eight_chain_model}b
plot the normalized stress $\Piola/\left(\refchains\forcemax\lenref\right)$
(Eq. \ref{eq:eight_chain_uniaxial}) and the damage $\damage=\pchain$
(Eq. \ref{eq:damage_general}) as a function of the stretch $\stretch$
for a polymer network. Since the chains are characterized by a limiting
force, the normalization $\refchains\forcemax\lenref=\beta_{m}\refchains\boltzmann T$,
where $\beta_{m}=\forcemax\lenref/\boltzmann T$ denotes the maximum
value of $\beta$ at the rupture force $\forcemax$, is used for convenience.
The proposed model recovers the classical eight-chain under small
to moderate stretches. However, at large deformations the proposed
model captures the extensibility of the repeat units, chain scission,
and the reduction in chains. Examination of the damage in \ref{fig:eight_chain_model}b
shows that at a stretch $\stretch\sim6.2$ chains begin to rupture
and the network fails. 

\section{Conclusions}

This contribution investigates the origin of failure in polymer chains.
To this end, we consider FJCs with extensible segments to account
for the entropic and the enthalpic energies that are stored in a chain
that is subjected to stretching. As opposed to other works, which
consider a variation in the stretching of the Kuhn segments through
the partition function, the proposed framework computes the Kuhn length
directly through the maximization of the entropy. The model reveals
that the external force applied to a chain is unequally distributed
along the repeat units. Specifically, segments that are aligned along
the end-to-end direction experience the maximum force. This allows
to compute the local probability of bond rupture, which is the highest
in repeat units that are aligned along the end-to-end vector. In turn,
one can determine the probability of chain scission. 

To capture the enthalpic contribution, a tilted Morse-potential is
employed to the Kuhn segments. Following previous works, we assume
that the probability of rupture of bonds along a Kuhn segment depend
on the activation energy. In the absence of a mechanical force, this
energy is high (on the order of $150\,\boltzmann T$) and therefore
bond rupture is highly unlikely. The application of an external force
shifts the energy landscape, and consequently two extrema points are
obtained along the potential energy - a minimum that describes the
equilibrium state and a maximum that characterizes the transition
state. The activation energy, which is defined as the difference between
these two energies, decreases with the applied force and accordingly
the probability of a rupture event increases. Once the activation
energy is zero, bond dissociation occurs spontaneously. Chain scission
occurs once one of the bonds break, and accordingly the probability
of failure can be computed. 

To demonstrate the merit of the framework, we examine the response
of chains with stiff C-C bonds and compliant physical bonds. The latter
is of particular interest to networks with sacrificial bonds or mechanophores,
in which weaker bonds can rupture prior to the main chain, interpenetrating
networks comprising two or more families of chains, where the failure
of specific chains activates other chains and provide a network with
enhanced toughness, and biological polymers in which various potential
failure mechanisms can be triggered. To enable predictive capabilities
under a general loading, we also show that the local chain model can
be integrated to continuum-level model representation of 3-dimensional
networks using the numerical micro-sphere method as well as the well-known
eight chain model. To account for the damage, we define a damage parameter
that is related to the gradual rupture of the chains. While we consider
monotonic uniaxial loading, the framework can be employed to capture
the response and the rupture of chains under cyclic loading. In this
case, one must note that the damage parameter can only increase and
therefore remains constant during the unloading phase. 

This work provides a physically-based model to quantify the stretching
and failure of a single chain and can be used to develop coarse grain
description of damage in polymer networks. 

\appendix

\section*{Appendix}

\section{Derivation of the force on a chain \label{sec:Derivation-of-force_on_chain}}

To derive the stretching force $\force$, we first write
\begin{equation}
\energyC-T\entropyC=\boltzmann T\left(-\monomers\ln\monomers+\alpha\monomers+\frac{\beta}{\lenref}\etoe\right),
\end{equation}
where the relation $\gamma=-1/\boltzmann T$ is employed. Note that
$\alpha=\alpha\left(\etoe\right)$ and $\beta=\beta\left(\etoe\right)$,
and therefore 
\begin{equation}
\frac{\partial\left(\energyC-T\entropyC\right)}{\partial\etoe}=\boltzmann T\left(\monomers\frac{\partial\alpha}{\partial\etoe}+\frac{\etoe}{\lenref}\frac{\partial\beta}{\partial\etoe}\right)+\frac{\boltzmann T}{\lenref}\beta.\label{eq:s2}
\end{equation}

We continue by calculating the derivatives. The Lagrange multiplier
is $\alpha=\ln\left(\monomers/\partition\right)$ and its derivative
is
\begin{equation}
\frac{\partial\alpha}{\partial\etoe}=-\frac{1}{\partition}\frac{\partial\partition}{\partial\beta}\frac{\partial\beta}{\partial\etoe}.\label{eq:s3}
\end{equation}
Substitution of Eq. \ref{eq:s3} into Eq. \ref{eq:s2} yields
\begin{equation}
\frac{\partial\left(\energyC-T\entropyC\right)}{\partial\etoe}=\boltzmann T\left(-\frac{\monomers}{\partition}\frac{\partial\partition}{\partial\beta}+\frac{\etoe}{\lenref}\right)\frac{\partial\beta}{\partial\etoe}+\frac{\boltzmann T}{\lenref}\beta.\label{eq:S4}
\end{equation}
The derivative of the partition function is determined from Eq. 11
in the main manuscript, 
\begin{align}
\frac{\monomers}{\partition}\frac{\partial\partition}{\partial\beta} & =\monomers\int_{0}^{\pi}\left(\frac{\lencur}{\lenref}\cos\theta+\left(\beta\frac{\cos\theta}{\lenref}-\frac{1}{\boltzmann T}\frac{\partial\energymon}{\partial\lencur}\right)\frac{\partial\lencur}{\partial\beta}\right)\pdf\solidangle\nonumber \\
 & =\monomers\int_{0}^{\pi}\frac{\lencur}{\lenref}\cos\theta\pdf\solidangle=\frac{\etoe}{\lenref},\label{eq:S5}
\end{align}
where we employ the relation $\partial\energymon/\partial\lencur=\boltzmann T\beta\cos\theta/\lenref$
(Eq. 7) and Eq. 13 in the main manuscript in passing. 

Finally, substitution of Eq. \ref{eq:S5} into Eq. \ref{eq:S4} yields
\begin{equation}
\force=\frac{\partial\left(\energyC-T\entropyC\right)}{\partial\etoe}=\frac{\boltzmann T}{\lenref}\beta.
\end{equation}

\section{The micro-sphere technique \label{sec:microsphere}}

To integrate from the chain to the network level, we employ the micro-sphere
technique proposed by \citet{Bazant&oh86ZAMM}. This method was later
used to capture the response on rubbery networks \citep{Miehe2004,cohen&mcmeeking19JMPS}. 

Consider a unit sphere with a surface that represents the end-to-end
directions of the chains. The directional averaging of a quantity
$\bullet$ can be approximated through the discrete summation 
\begin{equation}
\left\langle \bullet\right\rangle =\frac{1}{4\pi}\int_{A}\bullet\,\d A=\sum_{i=1}^{m}\bullet^{\left(i\right)}w^{\left(i\right)},
\end{equation}
where $i=1,...,m$ are the representative chain directions $\RTdir^{\left(i\right)}$,
$\bullet^{\left(i\right)}$ is the value of the variable being averaged
along the $i$-th direction, and $w^{\left(i\right)}$ are appropriate
non-negative weights, constrained by $\sum_{i=1}^{m}w^{\left(i\right)}=1$. 

In the special case of randomly oriented and uniformly distributed
networks, one requires $\sum_{i=1}^{m}\RTdir^{\left(i\right)}w^{\left(i\right)}=\mathbf{0}$
and $\sum_{i=1}^{m}\RTdir^{\left(i\right)}\otimes\RTdir^{\left(i\right)}w^{\left(i\right)}=1/3\mathbf{I}$. 

\citet{Bazant&oh86ZAMM} showed that a specific choice of $m=42$
representative directions provide sufficient accuracy for isotropically
distributed directions. In this work, we follow this conclusion. The
integration directions and the corresponding weight function are given
in Table 1 of \citet{Bazant&oh86ZAMM}.

\bibliographystyle{biochem}
\bibliography{chains}

@Article{cohen&mcmeeking19JMPS,
  author    = {Cohen, Noy and McMeeking, Robert M},
  journal   = {Journal of the Mechanics and Physics of Solids},
  title     = {On the swelling induced microstructural evolution of polymer networks in gels},
  year      = {2019},
  pages     = {666--680},
  volume    = {125},
  publisher = {Elsevier},
}

@Article{Zhang&etal00JPCB,
  author    = {Zhang, Wenke and Zou, Shan and Wang, Chi and Zhang, Xi},
  journal   = {The Journal of Physical Chemistry B},
  title     = {Single polymer chain elongation of poly (N-isopropylacrylamide) and poly (acrylamide) by atomic force microscopy},
  year      = {2000},
  number    = {44},
  pages     = {10258--10264},
  volume    = {104},
  publisher = {ACS Publications},
}

@Article{Chen&etal05macromol,
  author    = {Chen, Hongwei and Li, Junfang and Ding, Yanwei and Zhang, Guangzhao and Zhang, Qijin and Wu, Chi},
  journal   = {Macromolecules},
  title     = {Folding and unfolding of individual PNIPAM-g-PEO copolymer chains in dilute aqueous solutions},
  year      = {2005},
  number    = {10},
  pages     = {4403--4408},
  volume    = {38},
  publisher = {ACS Publications},
}

@Article{Treloar75book,
  author    = {Treloar, LR G},
  title     = {The physics of rubber elasticity},
  year      = {1975},
  publisher = {OUP Oxford},
}

@Book{Flory53book,
  author    = {Flory, Paul J},
  publisher = {Cornell university press},
  title     = {Principles of polymer chemistry},
  year      = {1953},
}

@Article{Bazant&oh86ZAMM,
  author    = {Ba{\v{z}}ant, P and Oh, BH838076},
  journal   = {ZAMM-Journal of Applied Mathematics and Mechanics/Zeitschrift f{\"u}r Angewandte Mathematik und Mechanik},
  title     = {Efficient numerical integration on the surface of a sphere},
  year      = {1986},
  number    = {1},
  pages     = {37--49},
  volume    = {66},
  publisher = {Wiley Online Library},
}

@Article{Kuhn1942,
  author    = {Kuhn, Werner and Grun, F},
  journal   = {Kolloid-Zeitschrift},
  title     = {Beziehungen zwischen elastischen Konstanten und Dehnungsdoppelbrechung hochelastischer Stoffe},
  year      = {1942},
  number    = {3},
  pages     = {248--271},
  volume    = {101},
  publisher = {Springer},
}

@Article{Miehe2004,
  author    = {Miehe, C and G{\"o}ktepe, Serdar and Lulei, F20965751091},
  journal   = {Journal of the Mechanics and Physics of Solids},
  title     = {A micro-macro approach to rubber-like materials-part I: the non-affine micro-sphere model of rubber elasticity},
  year      = {2004},
  number    = {11},
  pages     = {2617--2660},
  volume    = {52},
  publisher = {Elsevier},
}

@Article{Olive2024,
  author   = {Olive, Renata and Cohen, Noy},
  journal  = {Journal of the Mechanics and Physics of Solids},
  title    = {Deformation and failure mechanisms in spider silk fibers},
  year     = {2024},
  issn     = {0022-5096},
  pages    = {105480},
  volume   = {182},
  doi      = {10.1016/j.jmps.2023.105480},
  url      = {https://www.sciencedirect.com/science/article/pii/S0022509623002843},
}

@Article{Vernerey2018,
  author    = {Vernerey, Franck J and Brighenti, Roberto and Long, Rong and Shen, Tong},
  journal   = {Macromolecules},
  title     = {Statistical damage mechanics of polymer networks},
  year      = {2018},
  number    = {17},
  pages     = {6609--6622},
  volume    = {51},
  publisher = {ACS Publications},
}

@Article{Beyer2000,
  author   = {Beyer, Martin K.},
  journal  = {J. Chem. Phys.},
  title    = {The mechanical strength of a covalent bond calculated by density functional theory},
  year     = {2000},
  issn     = {0021-9606},
  month    = may,
  number   = {17},
  pages    = {7307--7312},
  volume   = {112},
  doi      = {10.1063/1.481330},
  url      = {https://doi.org/10.1063/1.481330},
}

@Article{Wang2019,
  author    = {Wang, Shu and Panyukov, Sergey and Rubinstein, Michael and Craig, Stephen L.},
  journal   = {Macromolecules},
  title     = {Quantitative Adjustment to the Molecular Energy Parameter in the Lake-Thomas Theory of Polymer Fracture Energy},
  year      = {2019},
  issn      = {0024-9297},
  month     = apr,
  number    = {7},
  pages     = {2772--2777},
  volume    = {52},
  comment   = {doi: 10.1021/acs.macromol.8b02341},
  doi       = {10.1021/acs.macromol.8b02341},
  publisher = {American Chemical Society},
  url       = {https://doi.org/10.1021/acs.macromol.8b02341},
}

@Article{Wang2002,
  author    = {Wang, Chi and Shi, Weiqing and Zhang, Wenke and Zhang, Xi and Katsumoto, Yukiteru and Ozaki, Yukihiro},
  journal   = {Nano Lett.},
  title     = {Force Spectroscopy Study on Poly(acrylamide) Derivatives: Effects of Substitutes and Buffers on Single-Chain Elasticity},
  year      = {2002},
  issn      = {1530-6984},
  month     = oct,
  number    = {10},
  pages     = {1169--1172},
  volume    = {2},
  comment   = {doi: 10.1021/nl0256917},
  doi       = {10.1021/nl0256917},
  publisher = {American Chemical Society},
  url       = {https://doi.org/10.1021/nl0256917},
}

@Article{Zhang2003,
  author   = {Zhang, Wenke and Zhang, Xi},
  journal  = {Progress in Polymer Science},
  title    = {Single molecule mechanochemistry of macromolecules},
  year     = {2003},
  issn     = {0079-6700},
  number   = {8},
  pages    = {1271--1295},
  volume   = {28},
  doi      = {10.1016/S0079-6700(03)00046-7},
  url      = {https://www.sciencedirect.com/science/article/pii/S0079670003000467},
}

@Article{Smith1996,
  author    = {Smith, Steven B. and Cui, Yujia and Bustamante, Carlos},
  journal   = {Science},
  title     = {Overstretching B-DNA: The Elastic Response of Individual Double-Stranded and Single-Stranded DNA Molecules},
  year      = {1996},
  month     = feb,
  number    = {5250},
  pages     = {795--799},
  volume    = {271},
  comment   = {doi: 10.1126/science.271.5250.795},
  doi       = {10.1126/science.271.5250.795},
  publisher = {American Association for the Advancement of Science},
  url       = {https://doi.org/10.1126/science.271.5250.795},
}

@Article{Ghatak2000,
  author    = {Ghatak, Animangsu and Vorvolakos, Katherine and She, Hongquan and Malotky, David L. and Chaudhury, Manoj K.},
  journal   = {J. Phys. Chem. B},
  title     = {Interfacial Rate Processes in Adhesion and Friction},
  year      = {2000},
  issn      = {1520-6106},
  month     = may,
  number    = {17},
  pages     = {4018--4030},
  volume    = {104},
  comment   = {doi: 10.1021/jp9942973},
  doi       = {10.1021/jp9942973},
  publisher = {American Chemical Society},
  url       = {https://doi.org/10.1021/jp9942973},
}

@Article{Wang2025,
  author   = {Wang, Shi-Qing and Fan, Zehao and Siavoshani, Asal and Wang, Ming-chi and Wang, Junpeng},
  journal  = {Extreme Mechanics Letters},
  title    = {Fresh considerations regarding time-dependent elastomeric fracture},
  year     = {2025},
  issn     = {2352-4316},
  pages    = {102277},
  volume   = {74},
  doi      = {10.1016/j.eml.2024.102277},
  url      = {https://www.sciencedirect.com/science/article/pii/S2352431624001573},
}

@Article{Lavoie2016,
  author   = {Lavoie, Shawn R. and Long, Rong and Tang, Tian},
  journal  = {Extreme Mechanics Letters},
  title    = {A rate-dependent damage model for elastomers at large strain},
  year     = {2016},
  issn     = {2352-4316},
  pages    = {114--124},
  volume   = {8},
  doi      = {10.1016/j.eml.2016.05.016},
  url      = {https://www.sciencedirect.com/science/article/pii/S2352431615300407},
}

@Article{Wang1952,
  author   = {Wang, Ming Chen and Guth, Eugene},
  journal  = {J. Chem. Phys.},
  title    = {Statistical Theory of Networks of Non-Gaussian Flexible Chains},
  year     = {1952},
  issn     = {0021-9606},
  month    = jul,
  number   = {7},
  pages    = {1144--1157},
  volume   = {20},
  doi      = {10.1063/1.1700682},
  url      = {https://doi.org/10.1063/1.1700682},
}

@Article{Treloar1946,
  author    = {Treloar, L. R. G.},
  journal   = {Trans. Faraday Soc.},
  title     = {The elasticity of a network of long-chain molecules.--III},
  year      = {1946},
  issn      = {0014-7672},
  number    = {0},
  pages     = {83--94},
  volume    = {42},
  doi       = {10.1039/TF9464200083},
  publisher = {The Royal Society of Chemistry},
  url       = {http://dx.doi.org/10.1039/TF9464200083},
}

@Article{Arruda1993,
  author   = {Arruda, Ellen M. and Boyce, Mary C.},
  journal  = {Journal of the Mechanics and Physics of Solids},
  title    = {A three-dimensional constitutive model for the large stretch behavior of rubber elastic materials},
  year     = {1993},
  issn     = {0022-5096},
  number   = {2},
  pages    = {389--412},
  volume   = {41},
  doi      = {10.1016/0022-5096(93)90013-6},
  url      = {https://www.sciencedirect.com/science/article/pii/0022509693900136},
}

@Article{Webber2007,
  author    = {Webber, Rebecca E. and Creton, Costantino and Brown, Hugh R. and Gong, Jian Ping},
  journal   = {Macromolecules},
  title     = {Large Strain Hysteresis and Mullins Effect of Tough Double-Network Hydrogels},
  year      = {2007},
  issn      = {0024-9297},
  month     = apr,
  number    = {8},
  pages     = {2919--2927},
  volume    = {40},
  comment   = {doi: 10.1021/ma062924y},
  doi       = {10.1021/ma062924y},
  publisher = {American Chemical Society},
  url       = {https://doi.org/10.1021/ma062924y},
}

@Article{Wang2011,
  author    = {Wang, Xiao and Hong, Wei},
  journal   = {Soft Matter},
  title     = {Pseudo-elasticity of a double network gel},
  year      = {2011},
  issn      = {1744-683X},
  number    = {18},
  pages     = {8576--8581},
  volume    = {7},
  doi       = {10.1039/C1SM05787A},
  publisher = {The Royal Society of Chemistry},
  url       = {http://dx.doi.org/10.1039/C1SM05787A},
}

@Article{Lavoie2019,
  author   = {Lavoie, Shawn R. and Millereau, Pierre and Creton, Costantino and Long, Rong and Tang, Tian},
  journal  = {Journal of the Mechanics and Physics of Solids},
  title    = {A continuum model for progressive damage in tough multinetwork elastomers},
  year     = {2019},
  issn     = {0022-5096},
  pages    = {523--549},
  volume   = {125},
  doi      = {10.1016/j.jmps.2019.01.001},
  url      = {https://www.sciencedirect.com/science/article/pii/S0022509618303740},
}

@Article{Yang2020,
  author   = {Yang, Tianhao and Liechti, Kenneth M. and Huang, Rui},
  journal  = {Journal of the Mechanics and Physics of Solids},
  title    = {A multiscale cohesive zone model for rate-dependent fracture of interfaces},
  year     = {2020},
  issn     = {0022-5096},
  pages    = {104142},
  volume   = {145},
  doi      = {10.1016/j.jmps.2020.104142},
  url      = {https://www.sciencedirect.com/science/article/pii/S0022509620303768},
}

@Article{Lake1967,
  author   = {Lake, G. J. and Thomas, A. G.},
  journal  = {Proc. A},
  title    = {The strength of highly elastic materials},
  year     = {1967},
  issn     = {0080-4630},
  month    = aug,
  number   = {1460},
  pages    = {108--119},
  volume   = {300},
  doi      = {10.1098/rspa.1967.0160},
  url      = {https://doi.org/10.1098/rspa.1967.0160},
}

@Article{Li2024,
  author   = {Li, Xueyu and Gong, Jian Ping},
  journal  = {Nature Reviews Materials},
  title    = {Design principles for strong and tough hydrogels},
  year     = {2024},
  issn     = {2058-8437},
  number   = {6},
  pages    = {380--398},
  volume   = {9},
  doi      = {10.1038/s41578-024-00672-3},
  refid    = {Li2024},
  url      = {https://doi.org/10.1038/s41578-024-00672-3},
}

@Article{Qian2017,
  author   = {Qian, Jin and Lin, Ji and Xu, Guang-Kui and Lin, Yuan and Gao, Huajian},
  journal  = {Journal of the Mechanics and Physics of Solids},
  title    = {Thermally assisted peeling of an elastic strip in adhesion with a substrate via molecular bonds},
  year     = {2017},
  issn     = {0022-5096},
  pages    = {197--208},
  volume   = {101},
  doi      = {10.1016/j.jmps.2017.01.007},
  url      = {https://www.sciencedirect.com/science/article/pii/S0022509616308298},
}

@Article{Keren2023,
  author   = {Keren, Shachar and Segal-Peretz, Tamar and Cohen, Noy},
  journal  = {Theoretical and Applied Fracture Mechanics},
  title    = {Exploiting perforations to enhance the adhesion of 3D-printed lap shears},
  year     = {2023},
  issn     = {0167-8442},
  pages    = {103986},
  volume   = {126},
  doi      = {10.1016/j.tafmec.2023.103986},
  url      = {https://www.sciencedirect.com/science/article/pii/S0167844223002495},
}

@Article{Du2011,
  author    = {Du, Ning and Yang, Zhen and Liu, Xiang Yang and Li, Yang and Xu, Hong Yao},
  journal   = {Adv. Funct. Mater.},
  title     = {Structural Origin of the Strain-Hardening of Spider Silk},
  year      = {2011},
  issn      = {1616-301X},
  month     = feb,
  number    = {4},
  pages     = {772--778},
  volume    = {21},
  doi       = {10.1002/adfm.201001397},
  publisher = {John Wiley & Sons, Ltd},
  url       = {https://doi.org/10.1002/adfm.201001397},
}

@Article{Cohen2025,
  author   = {Cohen, Noy and Zhang, Fuzhong},
  journal  = {Acta Biomaterialia},
  title    = {Modeling of protein networks reveals factors affecting stiffness, yield stress, and strain stiffening in silk fibers},
  year     = {2025},
  issn     = {1742-7061},
  pages    = {402--410},
  volume   = {208},
  doi      = {10.1016/j.actbio.2025.09.036},
  url      = {https://www.sciencedirect.com/science/article/pii/S1742706125007032},
}

@Article{Zhu2025,
  author    = {Zhu, Jie and Brassart, Laurence},
  journal   = {PRL},
  title     = {Stretching Response of a Polymer Chain with Deformable Bonds},
  year      = {2025},
  month     = may,
  number    = {21},
  pages     = {218101},
  volume    = {134},
  doi       = {10.1103/PhysRevLett.134.218101},
  publisher = {American Physical Society},
  refid     = {10.1103/PhysRevLett.134.218101},
  url       = {https://link.aps.org/doi/10.1103/PhysRevLett.134.218101},
}

@Article{Volokh2010,
  author   = {Volokh, K. Y.},
  journal  = {Mechanics Research Communications},
  title    = {On modeling failure of rubber-like materials},
  year     = {2010},
  issn     = {0093-6413},
  number   = {8},
  pages    = {684--689},
  volume   = {37},
  doi      = {10.1016/j.mechrescom.2010.10.006},
  url      = {https://www.sciencedirect.com/science/article/pii/S0093641310001461},
}

@Article{Volokh2013,
  author    = {Volokh, K. Y.},
  journal   = {Rubber Chemistry and Technology},
  title     = {Review of the energy limiters approach to modeling failure of rubber},
  year      = {2013},
  number    = {3},
  pages     = {470--487},
  volume    = {86},
  doi       = {10.5254/rct.13.87948},
  language  = {English},
  publisher = {The Rubber Division, American Chemical Society},
  url       = {https://rct.kglmeridian.com/view/journals/rcat/86/3/article-p470.xml},
}

@Article{Mulderrig2023,
  author   = {Mulderrig, Jason and Talamini, Brandon and Bouklas, Nikolaos},
  journal  = {Journal of the Mechanics and Physics of Solids},
  title    = {A statistical mechanics framework for polymer chain scission, based on the concepts of distorted bond potential and asymptotic matching},
  year     = {2023},
  issn     = {0022-5096},
  pages    = {105244},
  volume   = {174},
  doi      = {10.1016/j.jmps.2023.105244},
  url      = {https://www.sciencedirect.com/science/article/pii/S0022509623000480},
}

@Article{Evans1997,
  author   = {Evans, E. and Ritchie, K.},
  journal  = {Biophysical Journal},
  title    = {Dynamic strength of molecular adhesion bonds},
  year     = {1997},
  issn     = {0006-3495},
  number   = {4},
  pages    = {1541--1555},
  volume   = {72},
  doi      = {10.1016/S0006-3495(97)78802-7},
  url      = {https://www.sciencedirect.com/science/article/pii/S0006349597788027},
}

@Article{Kauzmann1940,
  author    = {Kauzmann, Walter and Eyring, Henry},
  journal   = {J. Am. Chem. Soc.},
  title     = {The Viscous Flow of Large Molecules},
  year      = {1940},
  issn      = {0002-7863},
  month     = nov,
  number    = {11},
  pages     = {3113--3125},
  volume    = {62},
  comment   = {doi: 10.1021/ja01868a059},
  doi       = {10.1021/ja01868a059},
  publisher = {American Chemical Society},
  url       = {https://doi.org/10.1021/ja01868a059},
}

@Article{Ducrot2014,
  author    = {Etienne Ducrot and Yulan Chen and Markus Bulters and Rint P. Sijbesma and Costantino Creton},
  journal   = {Science},
  title     = {Toughening Elastomers with Sacrificial Bonds and Watching Them Break},
  year      = {2014},
  month     = apr,
  number    = {6180},
  pages     = {186-189},
  volume    = {344},
  comment   = {doi: 10.1126/science.1248494},
  doi       = {10.1126/science.1248494},
  eprint    = {https://www.science.org/doi/pdf/10.1126/science.1248494},
  publisher = {American Association for the Advancement of Science},
  url       = {https://www.science.org/doi/abs/10.1126/science.1248494},
}

@Article{Buche2020,
  author    = {Buche, Michael R. and Silberstein, Meredith N.},
  journal   = {PRE},
  title     = {Statistical mechanical constitutive theory of polymer networks: The inextricable links between distribution, behavior, and ensemble},
  year      = {2020},
  month     = jul,
  number    = {1},
  pages     = {012501},
  volume    = {102},
  doi       = {10.1103/PhysRevE.102.012501},
  publisher = {American Physical Society},
  refid     = {10.1103/PhysRevE.102.012501},
  url       = {https://link.aps.org/doi/10.1103/PhysRevE.102.012501},
}

@Article{Buche2022,
  author    = {Buche, Michael R. and Silberstein, Meredith N. and Grutzik, Scott J.},
  journal   = {PRE},
  title     = {Freely jointed chain models with extensible links},
  year      = {2022},
  month     = aug,
  number    = {2},
  pages     = {024502},
  volume    = {106},
  doi       = {10.1103/PhysRevE.106.024502},
  publisher = {American Physical Society},
  refid     = {10.1103/PhysRevE.106.024502},
  url       = {https://link.aps.org/doi/10.1103/PhysRevE.106.024502},
}

@Article{Mulderrig2021,
  author   = {Mulderrig, Jason and Li, Bin and Bouklas, Nikolaos},
  journal  = {Mechanics of Materials},
  title    = {Affine and non-affine microsphere models for chain scission in polydisperse elastomer networks},
  year     = {2021},
  issn     = {0167-6636},
  pages    = {103857},
  volume   = {160},
  doi      = {10.1016/j.mechmat.2021.103857},
  url      = {https://www.sciencedirect.com/science/article/pii/S0167663621001083},
}

@Article{Buche2021,
  author   = {Buche, Michael R. and Silberstein, Meredith N.},
  journal  = {Journal of the Mechanics and Physics of Solids},
  title    = {Chain breaking in the statistical mechanical constitutive theory of polymer networks},
  year     = {2021},
  issn     = {0022-5096},
  pages    = {104593},
  volume   = {156},
  doi      = {10.1016/j.jmps.2021.104593},
  url      = {https://www.sciencedirect.com/science/article/pii/S0022509621002374},
}

@Article{Sheu2003,
  author    = {Sheu, Sheh-Yi and Yang, Dah-Yen and Selzle, H. L. and Schlag, E. W.},
  journal   = {Proceedings of the National Academy of Sciences},
  title     = {Energetics of hydrogen bonds in peptides},
  year      = {2003},
  month     = oct,
  number    = {22},
  pages     = {12683--12687},
  volume    = {100},
  comment   = {doi: 10.1073/pnas.2133366100},
  doi       = {10.1073/pnas.2133366100},
  publisher = {Proceedings of the National Academy of Sciences},
  url       = {https://doi.org/10.1073/pnas.2133366100},
}

@Article{Zhang2023,
  author    = {Zhang, Hang and Diesendruck, Charles E.},
  journal   = {Angew. Chem. Int. Ed.},
  title     = {Off-center Mechanophore Activation in Block Copolymers},
  year      = {2023},
  issn      = {1433-7851},
  month     = jan,
  number    = {2},
  pages     = {e202213980},
  volume    = {62},
  doi       = {10.1002/anie.202213980},
  publisher = {John Wiley & Sons, Ltd},
  url       = {https://doi.org/10.1002/anie.202213980},
}

@Article{Fantner2006,
  author   = {Fantner, Georg E. and Oroudjev, Emin and Schitter, Georg and Golde, Laura S. and Thurner, Philipp and Finch, Marquesa M. and Turner, Patricia and Gutsmann, Thomas and Morse, Daniel E. and Hansma, Helen and Hansma, Paul K.},
  journal  = {Biophysical Journal},
  title    = {Sacrificial Bonds and Hidden Length: Unraveling Molecular Mesostructures in Tough Materials},
  year     = {2006},
  issn     = {0006-3495},
  number   = {4},
  pages    = {1411--1418},
  volume   = {90},
  doi      = {10.1529/biophysj.105.069344},
  url      = {https://www.sciencedirect.com/science/article/pii/S0006349506723311},
}

@Article{Li2000,
  author    = {Li, Hongbin and Oberhauser, Andres F. and Fowler, Susan B. and Clarke, Jane and Fernandez, Julio M.},
  journal   = {Proceedings of the National Academy of Sciences},
  title     = {Atomic force microscopy reveals the mechanical design of a modular protein},
  year      = {2000},
  month     = jun,
  number    = {12},
  pages     = {6527--6531},
  volume    = {97},
  comment   = {doi: 10.1073/pnas.120048697},
  doi       = {10.1073/pnas.120048697},
  publisher = {Proceedings of the National Academy of Sciences},
  url       = {https://doi.org/10.1073/pnas.120048697},
}

@Article{Krautbauer2003,
  author    = {Krautbauer, Rupert and Rief, Matthias and Gaub, Hermann E.},
  journal   = {Nano Lett.},
  title     = {Unzipping DNA Oligomers},
  year      = {2003},
  issn      = {1530-6984},
  month     = apr,
  number    = {4},
  pages     = {493--496},
  volume    = {3},
  comment   = {doi: 10.1021/nl034049p},
  doi       = {10.1021/nl034049p},
  publisher = {American Chemical Society},
  url       = {https://doi.org/10.1021/nl034049p},
}

@Article{Bockelmann2002,
  author   = {Bockelmann, U. and Thomen, Ph. and Essevaz-Roulet, B. and Viasnoff, V. and Heslot, F.},
  journal  = {Biophysical Journal},
  title    = {Unzipping DNA with Optical Tweezers: High Sequence Sensitivity and Force Flips},
  year     = {2002},
  issn     = {0006-3495},
  number   = {3},
  pages    = {1537--1553},
  volume   = {82},
  doi      = {10.1016/S0006-3495(02)75506-9},
  url      = {https://www.sciencedirect.com/science/article/pii/S0006349502755069},
}

@Article{Cohen2021a,
  author    = {Cohen, Noy and Du, Cong and Wu, Zi Liang},
  journal   = {Macromolecules},
  title     = {Understanding the Dissociation of Hydrogen Bond Based Cross-Links In Hydrogels Due to Hydration and Mechanical Forces},
  year      = {2021},
  issn      = {0024-9297},
  month     = dec,
  pages     = {10.1021/acs.macromol.1c01927},
  comment   = {doi: 10.1021/acs.macromol.1c01927},
  doi       = {10.1021/acs.macromol.1c01927},
  publisher = {American Chemical Society},
  url       = {https://doi.org/10.1021/acs.macromol.1c01927},
}

@Article{Gong2010,
  author    = {Gong, Jian Ping},
  journal   = {Soft Matter},
  title     = {Why are double network hydrogels so tough?},
  year      = {2010},
  issn      = {1744-683X},
  number    = {12},
  pages     = {2583--2590},
  volume    = {6},
  doi       = {10.1039/B924290B},
  publisher = {The Royal Society of Chemistry},
  url       = {http://dx.doi.org/10.1039/B924290B},
}

@Article{Gong2003,
  author    = {Gong, J. P. and Katsuyama, Y. and Kurokawa, T. and Osada, Y.},
  journal   = {Adv. Mater.},
  title     = {Double-Network Hydrogels with Extremely High Mechanical Strength},
  year      = {2003},
  issn      = {0935-9648},
  month     = jul,
  number    = {14},
  pages     = {1155--1158},
  volume    = {15},
  doi       = {10.1002/adma.200304907},
  publisher = {John Wiley & Sons, Ltd},
  url       = {https://doi.org/10.1002/adma.200304907},
}

@Article{Olive2025,
  author   = {Olive, Renata and Cohen, Noy},
  journal  = {International Journal of Solids and Structures},
  title    = {Employing spinning conditions to control the mechanical response of spider silk fibers},
  year     = {2025},
  issn     = {0020-7683},
  pages    = {113592},
  volume   = {322},
  doi      = {10.1016/j.ijsolstr.2025.113592},
  url      = {https://www.sciencedirect.com/science/article/pii/S0020768325003786},
}

@Article{Itskov2016,
  author   = {Itskov, Mikhail and Knyazeva, Anna},
  journal  = {International Journal of Solids and Structures},
  title    = {A rubber elasticity and softening model based on chain length statistics},
  year     = {2016},
  issn     = {0020-7683},
  pages    = {512--519},
  volume   = {80},
  doi      = {10.1016/j.ijsolstr.2015.10.011},
  url      = {https://www.sciencedirect.com/science/article/pii/S0020768315004278},
}

@Article{Hillgaertner2018,
  author   = {Hillgartner, Markus and Linka, Kevin and Itskov, Mikhail},
  journal  = {Journal of Biomechanics},
  title    = {Worm-like chain model extensions for highly stretched tropocollagen molecules},
  year     = {2018},
  issn     = {0021-9290},
  pages    = {129--135},
  volume   = {80},
  doi      = {10.1016/j.jbiomech.2018.08.034},
  url      = {https://www.sciencedirect.com/science/article/pii/S002192901830719X},
}

@Article{Itskov2016a,
  author   = {Itskov, Mikhail and Darabi, Ehsan},
  journal  = {Composites Part B: Engineering},
  title    = {Constitutive modeling of carbon nanotube rubber composites on the basis of chain length statistics},
  year     = {2016},
  issn     = {1359-8368},
  pages    = {69--75},
  volume   = {90},
  doi      = {10.1016/j.compositesb.2015.12.002},
  url      = {https://www.sciencedirect.com/science/article/pii/S1359836815007258},
}

@Article{Lamont2021,
  author    = {Lamont, Samuel C. and Mulderrig, Jason and Bouklas, Nikolaos and Vernerey, Franck J.},
  journal   = {Macromolecules},
  title     = {Rate-Dependent Damage Mechanics of Polymer Networks with Reversible Bonds},
  year      = {2021},
  issn      = {0024-9297},
  month     = dec,
  number    = {23},
  pages     = {10801--10813},
  volume    = {54},
  comment   = {doi: 10.1021/acs.macromol.1c01943},
  doi       = {10.1021/acs.macromol.1c01943},
  publisher = {American Chemical Society},
  url       = {https://doi.org/10.1021/acs.macromol.1c01943},
}

@Article{Lamont2025,
  author   = {Lamont, Samuel C. and Bouklas, Nikolaos and Vernerey, Franck J.},
  journal  = {International Journal of Fracture},
  title    = {Cohesive instability in elastomers: insights from a crosslinked Van der Waals fluid model},
  year     = {2025},
  issn     = {1573-2673},
  number   = {1},
  pages    = {20},
  volume   = {249},
  doi      = {10.1007/s10704-025-00840-8},
  refid    = {Lamont2025},
  url      = {https://doi.org/10.1007/s10704-025-00840-8},
}

@Article{Mao2017,
  author   = {Mao, Yunwei and Talamini, Brandon and Anand, Lallit},
  journal  = {Extreme Mechanics Letters},
  title    = {Rupture of polymers by chain scission},
  year     = {2017},
  issn     = {2352-4316},
  pages    = {17--24},
  volume   = {13},
  doi      = {10.1016/j.eml.2017.01.003},
  url      = {https://www.sciencedirect.com/science/article/pii/S2352431616302681},
}

@Article{Brown1994,
  author    = {Brown, Hugh R. and Hui, Chung Yuen and Raphael, Elie},
  journal   = {Macromolecules},
  title     = {Interplay between intermolecular interactions and chain pullout in the adhesion of elastomer},
  year      = {1994},
  issn      = {0024-9297},
  month     = jan,
  number    = {2},
  pages     = {608--609},
  volume    = {27},
  comment   = {doi: 10.1021/ma00080a041},
  doi       = {10.1021/ma00080a041},
  publisher = {American Chemical Society},
  url       = {https://doi.org/10.1021/ma00080a041},
}

@InCollection{Creton2002,
  author    = {Creton, Costantino and Kramer, Edward J. and Brown, Hugh R. and Hui, Chung-Yuen},
  booktitle = {Molecular Simulation Fracture Gel Theory},
  publisher = {Springer Berlin Heidelberg},
  title     = {Adhesion and Fracture of Interfaces Between Immiscible Polymers: from the Molecular to the Continuum Scal},
  year      = {2002},
  address   = {Berlin, Heidelberg},
  pages     = {53--136},
  doi       = {10.1007/3-540-45141-2_2},
  issn      = {978-3-540-45141-9},
  refid     = {Creton2002},
  url       = {https://doi.org/10.1007/3-540-45141-2_2},
}

@Article{Slootman2020,
  author    = {Slootman, Juliette and Waltz, Victoria and Yeh, C. Joshua and Baumann, Christoph and G\"ostl, Robert and Comtet, Jean and Creton, Costantino},
  journal   = {Phys. Rev. X},
  title     = {Quantifying Rate- and Temperature-Dependent Molecular Damage in Elastomer Fracture},
  year      = {2020},
  month     = {Dec},
  pages     = {041045},
  volume    = {10},
  doi       = {10.1103/PhysRevX.10.041045},
  issue     = {4},
  numpages  = {14},
  publisher = {American Physical Society},
  url       = {https://link.aps.org/doi/10.1103/PhysRevX.10.041045},
}

@Article{Miehe2014,
  author   = {Miehe, Christian and Schanzel, Lisa-Marie},
  journal  = {Journal of the Mechanics and Physics of Solids},
  title    = {Phase field modeling of fracture in rubbery polymers. Part I: Finite elasticity coupled with brittle failure},
  year     = {2014},
  issn     = {0022-5096},
  pages    = {93--113},
  volume   = {65},
  doi      = {10.1016/j.jmps.2013.06.007},
  url      = {https://www.sciencedirect.com/science/article/pii/S0022509613001191},
}

@Article{Ang2022,
  author    = {Ang, Ida and Bouklas, Nikolaos and Li, Bin},
  journal   = {Int J Numer Methods Eng},
  title     = {Stabilized formulation for phase-field fracture in nearly incompressible hyperelasticity},
  year      = {2022},
  issn      = {0029-5981},
  month     = oct,
  number    = {19},
  pages     = {4655--4673},
  volume    = {123},
  doi       = {10.1002/nme.7050},
  publisher = {John Wiley & Sons, Ltd},
  url       = {https://doi.org/10.1002/nme.7050},
}

@Article{PEERLINGS1996,
  author    = {Peerlings, R. H. J. and De Borst, R. and Brekelmans, W. A. M. and De Vree, J. H. P.},
  journal   = {Int. J. Numer. Meth. Engng.},
  title     = {Gradient enhanced damage for quasi-brittle materials},
  year      = {1996},
  issn      = {0029-5981},
  month     = oct,
  number    = {19},
  pages     = {3391--3403},
  volume    = {39},
  doi       = {10.1002/(SICI)1097-0207(19961015)39:19<3391::AID-NME7>3.0.CO;2-D},
  publisher = {John Wiley & Sons, Ltd},
  url       = {https://doi.org/10.1002/(SICI)1097-0207(19961015)39:19<3391::AID-NME7>3.0.CO;2-D},
}

@Article{Mousavi2024,
  author   = {Mousavi, S. Mohammad and Ang, Ida and Mulderrig, Jason and Bouklas, Nikolaos},
  journal  = {J. Appl. Mech},
  title    = {Evaluating Fracture Energy Predictions Using Phase-Field and Gradient-Enhanced Damage Models for Elastomers},
  year     = {2024},
  issn     = {0021-8936},
  month    = oct,
  number   = {12},
  volume   = {91},
  doi      = {10.1115/1.4066385},
  url      = {https://doi.org/10.1115/1.4066385},
}

@Article{Talamini2018,
  author   = {Talamini, Brandon and Mao, Yunwei and Anand, Lallit},
  journal  = {Journal of the Mechanics and Physics of Solids},
  title    = {Progressive damage and rupture in polymers},
  year     = {2018},
  issn     = {0022-5096},
  pages    = {434--457},
  volume   = {111},
  doi      = {10.1016/j.jmps.2017.11.013},
  url      = {https://www.sciencedirect.com/science/article/pii/S0022509617303939},
}

@Article{Mousavi2025,
  author   = {Mousavi, S. Mohammad and Mulderrig, Jason and Talamini, Brandon and Bouklas, Nikolaos},
  journal  = {Computer Methods in Applied Mechanics and Engineering},
  title    = {A chain stretch-based gradient-enhanced model for damage and fracture in elastomers},
  year     = {2025},
  issn     = {0045-7825},
  pages    = {118103},
  volume   = {444},
  doi      = {10.1016/j.cma.2025.118103},
  url      = {https://www.sciencedirect.com/science/article/pii/S0045782525003755},
}

@Article{Mousavi2026,
  author   = {Mousavi, S. Mohammad and Mulderrig, Jason and Talamini, Brandon and Bouklas, Nikolaos},
  journal  = {Journal of the Mechanics and Physics of Solids},
  title    = {Capturing the fractocohesive length scale in elastomers through a statistical mechanics-based gradient enhanced damage model},
  year     = {2026},
  issn     = {0022-5096},
  pages    = {106504},
  volume   = {209},
  doi      = {10.1016/j.jmps.2026.106504},
  url      = {https://www.sciencedirect.com/science/article/pii/S0022509626000049},
}

@Article{Cohen2019a,
  author   = {Cohen, Noy},
  journal  = {International Journal of Solids and Structures},
  title    = {Programming the equilibrium swelling response of heterogeneous polymeric gels},
  year     = {2019},
  issn     = {0020-7683},
  pages    = {81--90},
  volume   = {178-179},
  doi      = {10.1016/j.ijsolstr.2019.06.023},
  url      = {https://www.sciencedirect.com/science/article/pii/S0020768319303063},
}

@Article{Abkenar2017,
  author    = {Abkenar, Masoud and Gray, Thomas H. and Zaccone, Alessio},
  journal   = {PRE},
  title     = {Dissociation rates from single-molecule pulling experiments under large thermal fluctuations or large applied force},
  year      = {2017},
  month     = apr,
  number    = {4},
  pages     = {042413},
  volume    = {95},
  doi       = {10.1103/PhysRevE.95.042413},
  publisher = {American Physical Society},
  refid     = {10.1103/PhysRevE.95.042413},
  url       = {https://link.aps.org/doi/10.1103/PhysRevE.95.042413},
}

@Article{Cohen2026,
  author    = {Cohen, Noy},
  journal   = {Macromolecules},
  title     = {Phase Coexistence in Thermoresponsive PNIPAM Hydrogels Triggered by Mechanical Forces},
  year      = {2026},
  issn      = {0024-9297},
  month     = feb,
  number    = {4},
  pages     = {2580--2589},
  volume    = {59},
  comment   = {doi: 10.1021/acs.macromol.5c03088},
  doi       = {10.1021/acs.macromol.5c03088},
  publisher = {American Chemical Society},
  url       = {https://doi.org/10.1021/acs.macromol.5c03088},
}

\end{document}